\documentclass[final,authoryear,3p,times,10pt]{elsarticle}

\usepackage{multirow,setspace,times,amssymb,amsmath,graphicx,color,rotating,subfigure,url}
\usepackage{lineno,color}
\usepackage{natbib}
\usepackage{booktabs}%
\usepackage{longtable}%
\usepackage{rotating}
\usepackage{lscape}
\usepackage{ulem}
\graphicspath{{Figures/}}
\usepackage[table]{xcolor}
\usepackage{tabularx}
\usepackage{graphicx} %use graph format
\usepackage{epstopdf}
\usepackage{mathrsfs}
\usepackage{makecell}
\usepackage[bookmarks=true,colorlinks,linkcolor=blue,anchorcolor=blue,citecolor=blue,unicode]{hyperref}
\usepackage{bookmark}

\hypersetup{CJKbookmarks=true}%

\journal{Research in International Business and Finance}

\begin{document}
% \begin{CJK*}{GBK}{Song} % Use default fonts from CJK (see below)
\newcommand{\Rmnum}[1]{\uppercase\expandafter{\romannumeral #1}}  

\begin{frontmatter}
%\newpage
\title{Correlation structure analysis of the global agricultural futures market}

\author[SB]{Yun-Shi Dai}
\author[VN]{Ngoc Quang Anh Huynh}
\author[SB,RCE]{Qing-Huan Zheng}
% % \ead{wjxie@ecust.edu.cn}
\author[SB,RCE,DM]{Wei-Xing Zhou\corref{WXZ}}
\ead{wxzhou@ecust.edu.cn}
\cortext[WXZ]{Corresponding author.} %Corresponding to: 130 Meilong Road, P.O. Box 114, School of Business, East China University of Science and Technology, Shanghai 200237, China.}

\address[SB]{School of Business, East China University of Science and Technology, Shanghai, China}
\address[VN]{College of Technology and Design, University of Economics Ho Chi Minh City (UEH University), Ho Chi Minh City, Vietnam}
\address[RCE]{Research Center for Econophysics, East China University of Science and Technology, Shanghai, China}
\address[DM]{School of Mathematics, East China University of Science and Technology, Shanghai, China}

\begin{abstract}
This paper adopts the random matrix theory (RMT) to analyze the correlation structure of the global agricultural futures market from 2000 to 2020. It is found that the distribution of correlation coefficients is asymmetric and right skewed, and many eigenvalues of the correlation matrix deviate from the RMT prediction. The largest eigenvalue reflects a collective market effect common to all agricultural futures, the other largest deviating eigenvalues can be implemented to identify futures groups, and there are modular structures based on regional properties or agricultural commodities among the significant participants of their corresponding eigenvectors. Except for the smallest eigenvalue, other smallest deviating eigenvalues represent the agricultural futures pairs with highest correlations. This paper can be of reference and significance for using agricultural futures to manage risk and optimize asset allocation.
\end{abstract}


\end{frontmatter}

% \tableofcontents

\section{Introduction}

Food is a basic human need, but food security is always being challenged. Inadequate production and economic stagnation led to the world food crisis of 1972-1974, the African drought caused an unprecedented food crisis of 1982-1986, and a series of natural disasters, the overexploitation of biofuels and soaring food prices triggered the world food crisis of 2007-2008. In fact, all sorts of information shocks can exert influence on the prices of agricultural products and then lead to crises, because agricultural commodities are supplied by a large number of diverse and heterogeneous producers. According to the {\textit{Global Report on Food Crises 2021}} released by the Food and Agriculture Organization of the United Nations (FAO)\footnote{https://www.fao.org/resilience/resources/recursos-detalle/es/c/1398545/}, persistent conflict, weather extremes and economic shocks have been the three primary drivers of acute food insecurity. Meanwhile, the COVID-19 pandemic and various containment measures around the world have widened inequalities, aggravated situations and magnified vulnerabilities of the global food system.

Futures trading began in agricultural commodities. As the earliest type of futures in the world, agricultural futures have always been an indispensable part of the international futures market. With the rapid development of commodity economy and the great improvement of productivity, numerous agricultural futures markets have emerged all over the world, and the varieties of agricultural futures available for trade have gradually expanded from the initial grain futures to cash crop futures, livestock products futures, and wood products futures. According to the global exchange-traded derivatives data released by Futures Industry Association (FIA), the turnover of exchange-traded derivatives based on agricultural commodities reached 2.569 billion lots in 2020, with a year-on-year increase of 45.35\%, among which the volume of transactions in the Asia-Pacific region led the world.

% 2021/11/20, 18:55

Price discovery and risk transfer are two key functions of commodity futures markets. Futures markets can reveal information about forward prices in the spot market, and further mitigate the price risk related to commodities \citep{Joseph-Sisodia-Tiwari-2014-EconModel}. For various agricultural products, the majority of previous research argues that futures market plays a leading role in the process of price discovery \citep{Zapata-Fortenbery-1996-ApplEconPerspectPolicy, Yang-Bessler-Leatham-2001-JFuturesMark, Yang-Zhang-2013-EconModel}. The essence of price discovery is to provide a reference price so that the associated spot price can be discovered. Hence, agricultural futures price often act as a powerful predictor for the prospective price of spot market. In addition, hedging in agricultural futures markets enables producers or consumers to effectively avoid, transfer or disperse the risk of price fluctuations in the spot market. However, from another perspective, futures prices dominate spot prices, indicating that the systemic risk in the futures market is likely to spill over to the spot market, and then lead to a crisis in the spot market for agricultural commodities. Therefore, it is significant to evaluate the systemic risk of the agricultural futures market, and the most critical one is to clarify its correlation structure.

Diverse methods can be applied to recognize correlations in the financial system, such as minimum spanning tree (MST), planar maximally filtered graph (PMFG) and so on \citep{Naeem-Karim-2021-EconLett, Karim-Lucey-Naeem-Uddin-2022-FinancResLett}. However, the strengths of random matrix theory (RMT) are manifested clearly when it is used to clarify the correlation structure of a financial market. Specifically, the maximum eigenvalue of the correlation structure can reflect the collective behavior of the market, and other eigenvalues can identify the clustering groups with certain characteristics. Thus, the application of random matrix theory in agricultural futures markets merit further investigation, on which current academic research is still a blank. Making a conscious effort to fill this gap is the focus of our research. The main contribution of this paper resides in trying to understand the deviation from randomness as a source of dependencies, which may pose as systemic risk in the global agricultural futures market.

%contributions

We select 74 typical agricultural futures in different agricultural futures markets around the world, and use their daily closing prices of continuous contracts from 2000 to 2020 to construct the correlation matrix of returns. By comparing the statistical properties of the empirical correlation matrix with those of a random correlation matrix, we aim to test the randomness of the correlation matrix and distinguish the parts of it that deviate from the random matrix. We further reveal abundant economic information of the deviating eigenvalues and their corresponding eigenvectors, which suggests that the correlation structure of the global agricultural futures market is quite specific and characteristic. Specially, we construct an agricultural futures price index (AFPI) for the global agricultural futures market based on the eigenportfolio corresponding to the largest eigenvalue, which performs better than the average price series under the buy-and-hold strategy.

The rest of the paper is organized as follows. Section~\ref{S1:LitRev} is literature review. Section~\ref{S1:Data:Methodology} introduces the data sets and presents the statistical description. Section~\ref{S1:EmpAnal} uses the random matrix theory \sout{(RMT)} to analyze the empirical correlation structure of the global agricultural futures market, and investigates the economic information involved in the deviating eigenvalues and their corresponding eigenvectors of the empirical correlation matrix. Section~\ref{S1:Conclude} contains some conclusions and implications.

% 2021/12/05, 15:27

\section{Literature review} 
\label{S1:LitRev}

\subsection{Agricultural futures market}

In view of the important significance of agricultural futures markets, many scholars have carried out in-depth research on agricultural futures markets from theoretical level and empirical level respectively.

Since their inception, agricultural futures markets have always played an indispensable role in price discovery and risk diversification. \cite{McKenzie-Holt-2002-ApplEcon} tested the unbiasedness and efficiency of four different agricultural futures markets, indicating that some markets may show short-run pricing bias and inefficiency, but each futures market is unbiased in the long term. \cite{Dimpfl-Flad-Jung-2017-JCommodMark} investigated the relationship between spot prices and futures prices of eight agricultural commodities to test which market leads price discovery, and found evidence that the futures market contributes to price discovery limitedly and futures speculation will not distort commodity prices in the long term. \cite{Ke-Li-McKenzie-Liu-2019-Sustainability} examined the risk transfer between Chinese and the US agricultural futures markets with CoVaR, and confirmed that the US agricultural futures market dominates in price discovery while Chinese market performs an increasing role. \cite{Yang-Li-Wang-2021-JFuturesMark} selected 11 agricultural futures with the largest trading volumes in China and investigated their performance in price discovery. \cite{Sifat-Ghafoor-Mand-2021-JBehavExpFinanc} revealed diverse speculations in commodity futures during crisis periods, finding that volatility is closely and often non-linearly related to speculation, and agricultural futures bear greater hedging pressure while precious metal and energy futures are more prevalent.

With the global economic integration, the linkage between agricultural futures markets in different regions all over the world has become closer and closer. \cite{Li-Lu-2012-PhysicaA} examined the correlation between agricultural futures markets of the US and China, proving that the correlation between them is significantly multifractal, which is consistent with the empirical results of \cite{He-Chen-2011-ChaosSolitonsFractals}. By applying the thermal optimal path method (TOP), \cite{Jia-Wang-Tu-Li-2016-PhysicaA} selected three major agricultural products to explore the dynamic lead-lag relationships in both volatility and returns between the US and Chinese futures markets. \cite{Adammer-Bohl-vonLedebur-2017-BullEconRes} analyzed price dynamics of agricultural commodity futures in both short- and long-term between the US and European futures markets, and acknowledged that while the US market predominates in price transmissions and volatility spillovers, the impact of European futures market is rising on a global scale.

In addition, the interactions between agricultural markets and other markets have also been significantly enhanced with economic globalization and technical advancement. Over the past 20 years, a flood of literature has studied the relationship between agricultural commodities and crude oil \citep{Nazlioglu-Erdem-Soytas-2013-EnergyEcon,Wang-Wu-Yang-2014-EnergyEcon,Kang-McIver-Yoon-2017-EnergyEcon,Ji-Bouri-Roubaud-Shahzad-2018-EnergyEcon,Tiwari-Boachie-Suleman-Gupta-2021-Energy}. Specially, \cite{Naeem-Farid-Nor-Shahzad-2021-Mathematics,Naeem-Karim-Hasan-Kang-2022-SSRN} investigated the spillover network between agricultural commodities and oil shocks, and compared their dynamic relationships during the global financial crisis, the Shale Oil Revolution, and the COVID-19 episodes. Different asset classes are usually considered for risk management and portfolio diversification. Thus, \cite{Naeem-Hasan-Arif-Suleman-Kang-2022-EnergyEcon} further examined the safe-haven and hedging role of oil for agricultural commodities by applying quantile-on-quantile regression (QQR), and compared the hedge effectiveness before and after the global financial crisis. Apart from energy and metals markets, some literature also tries to connect agricultural markets with some emerging financial markets for novel analysis, such as the cryptocurrency market \citep{Naeem-Farid-Balli-Shahzad-2021-ApplEconLett}, the green bond market \citep{Nguyen-Naeem-Balli-Balli-Vo-2021-FinancResLett,Naeem-Adekoya-Oliyide-2021-JCleanProd} and the renewable energy equity market \citep{Alola-2022-Energy}.

A few scholars have attempted to investigate the correlation structures of different agricultural commodities. \cite{Boroumand-Goutte-Porcher-Porcher-2014-ApplEconLett} used estimated correlation matrix and principal component analysis (PCA) to study the correlation structures of a large quantity of agricultural commodities prices in France, demonstrating that there exist various degrees of correlations among different agricultural commodities and the price behaviors of some commodities have similar trends. Conversely, \cite{Sensoy-Hacihasanoglu-Nguyen-2015-ResourPolicy} found no empirical evidence of convergence among agricultural futures prices, and the majority of them moved in an uncorrelated way. \cite{Balli-Naeem-Shahzad-deBruin-2019-EnergyEcon} measured individual uncertainties of 22 commodities including agricultural commodities, and further described the connectedness and spillover network between different commodity classes. \cite{Xiao-Yu-Fang-Ding-2020-JFuturesMark} used a network method of variance decomposition to calculate the connectivity of 18 different types of commodity futures from static and dynamic perspectives, and showed that agricultural commodity futures are more vulnerable to the impact of other commodity futures and the connectivity generally increases in times of economic turmoil.

Analyzing the correlation between different agricultural futures has both theoretical value and practical significance, but there is little research about the correlation structure of the global agricultural futures market.

\subsection{Random matrix theory}

Since it was proposed in the 1930s, random matrix theory has always attracted much attention. In the 1950s, the discovery of Wigner's semicircle law \citep{Wigner-1955-AnnMath,Wigner-1958-AnnMath} further stimulated scholars' interest in the study of random matrices and high-dimensional random matrices. Mehta systematically expounded random matrix theory in his book, and gave an elaborate and comprehensive introduction to diverse methods for analyzing random matrices \citep{Mehta-1991}. Since then, random matrix theory has been widely implemented in physics, biology, communication and other fields to carry out academic research.

Financial market is a highly complex dynamic system \citep{Jiang-Xie-Zhou-Sornette-2019-RepProgPhys}, thus it is quite difficult to quantify the internal interactions of it. In the context, econophysics came into being, whose name first emerged in 1995 with \cite{Stanley-Afanasyev-Amaral-Buldyrev-Goldberger-Havlin-Leschhorn-Maass-Mantegna-Peng-Prince-Salinger-Stanley-Viswanathan-1996-PhysicaA} at a conference 
about complex systems held in Kolkata. It is a branch of complex system physics, which attempts to make use of the pre-existing large amount of data and the methods of statistical physics, to conduct a comprehensive investigation on statistical properties of financial markets \citep{Mantegna-Stanley-2000}. Scholars gradually find that there are deep relationships between physical system and financial system \citep{Sornette-2014-RepProgPhys, Huber-Sornette-2016-EPJst}, which also lays a foundation for applying random matrix theory in finance.

\cite{Laloux-Cizeau-Bouchaud-Potters-1999-PhysRevLett} constructed the correlation matrix based on daily variations of 406 stocks of the S\&P 500 during 1991–1996, and obtained the remarkable consistence between the theoretical prediction of RMT and the empirical density of eigenvalues associated to the time series of these different stocks, which confirms that there exists plenty of noise in the measured correlation matrix. \cite{Plerou-Gopikrishnan-Rosenow-Amaral-Stanley-2000-PhysicaA, Plerou-Gopikrishnan-Rosenow-Amaral-Guhr-Stanley-2002-PhysRevE} calculated multiple correlation matrices of stock returns, respectively, constructed by 30-minute data of 1000 stocks for the biennium 1994–1995, 30-minute data of 881 stocks for the biennium 1996–1997 and daily data of 422 stocks for the 35-year period 1962–1996. The analysis of statistical properties of these cross-correlation matrices showed that only a few minimum and maximum eigenvalues deviate from the prediction range of RMT, and their corresponding eigenvectors are stable in time.

In addition to taking the American stock market as the research object, some scholars also apply random matrix theory to analyze stock markets in other regions. \cite{Wilcox-Gebbie-2004-PhysicaA} empirically investigated correlations of the stock data from South African financial market, and found that compared with the American stock market, South African stock market has a higher proportion of eigenvalues deviating from theoretical prediction range. \cite{Garas-Argyrakis-2007-PhysicaA} selected three various portfolios which are traded in the Athens Stock Exchange during 1987–2004, and used both random matrix theory and minimum spanning tree to examine the evolution of these portfolios and the entire stock market simultaneously. \cite{Ren-Zhou-2014-PLoSOne} calculated multiple correlation matrices constructed from the returns of 367 Chinese stocks over a moving window, and brought a dynamic perspective to demonstrate the evolution of correlations in Chinese stock market. \cite{Han-Xie-Xiong-Zhang-Zhou-2017-FluctNoiseLett} applied random matrix theory to analyze and compare the statistical properties of Chinese stock market before and after the financial crisis of 2008.

Some literature discusses the role of random matrix theory in portfolio construction. \cite{Sharifi-Crane-Shamaie-Ruskin-2004-PhysicaA} sought to extract the non-noisy part of an empirical correlation matrix, and proposed a new technique of filtering tested by the Krzanowski model, which offers valuable insight for portfolio optimization. \cite{Daly-Crane-Ruskin-2008-PhysicaA} studied different effects of three RMT filters on the realized risk of asset portfolios, and on the stability of covariance matrices, finding that RMT filters reduce the realized risk on average, but increase the realized risk in some specific cases. \cite{Eom-Park-2018-Pac-BasinFinancJ} removed a market factor of the sample correlation matrix through RMT, and devised a new method to estimate the correlation matrix, which can be used to construct a diversified portfolio and gain better investment performance. \cite{Mo-Chen-2021-IEEEAccess} empirically proved that the RMT method can filter the risk measurement and make an improvement in the Markowitz optimization process to boost returns on portfolios.

A few scholars also pay attention to the application of random matrix theory in other fields closely related to finance, such as the housing market \citep{Meng-Xie-Jiang-Podobnik-Zhou-Stanley-2014-SciRep, Meng-Xie-Zhou-2015-IntJModPhysB}, the global crude oil market \citep{Dai-Xie-Jiang-Jiang-Zhou-2016-EmpirEcon} and the \sout{uncertainty of national} economic policy \citep{Ji-Huang-Cao-Hu-2019-ConcurrComput-PractExp, Dai-Xiong-Zhou-2021-FinancResLett}. However, to our knowledge, very few academic papers have focused on the application of random matrix theory in agricultural futures markets.

\section{Data and methodology}
\label{S1:Data:Methodology}

\subsection{Data sets}
 
The global agricultural futures market can be divided into the Asia Pacific market, the North American market, the Latin American market, and the European market. The Asia Pacific market mainly includes Australia, China, India, Malaysia, and New Zealand. In detail, Zhengzhou Commodity Exchange (CZCE) and Dalian Commodity Exchange (DCE) are the main trading places of agricultural futures in China. India's agricultural futures exchanges include Multi Commodity Exchange (MCX) and National Commodlity and Derivative Exchange (NCDEX). New Zealand, Australia and Malaysia have New Zealand Exchange (NZX), Sydney Futures Exchange (SFE) and Malaysia Derivatives Exchange (MDEX) respectively. The Chicago Board of Trade (CBOT), Chicago Mercantile Exchange (CME) and Intercontinental Exchange (ICE) are the main agricultural futures exchanges of the North American market. The Latin American market has BM\&FBOVESPA (BMF) and the European market has European Exchange (EUREX).

We select 74 agricultural futures traded on various exchanges around the world, and obtain the daily closing price series of their continuous contracts (2000–2020) from the Wind (https://www.wind.com.cn) and Bloomberg (https://www.bloomberg.com) databases. Considering the significant differences in the listing dates of different agricultural futures and the operation mechanism of different markets, we align the price time series according to the following steps. For the empty data before the listing of agricultural futures, we use their closing prices of the listing day as the surrogate. Because missing data points still exist, we further use the prices of the preceding days as a complement to them. In case that the data points of the previous day are still missing, we make up with the prices of two days ago, and the rest are complemented in the same way. The labels of different agricultural futures, their corresponding exchanges and countries are presented in Table \ref{Tab:AgroFutures:Stat}.

\subsection{Statistical description}

In order to quantify correlation structure of the global agricultural futures market, we first calculate the daily logarithmic return of the $i$-th agricultural futures over a time scale $\Delta t$ as follows:
\begin{equation}\label{Eq:Logarithmic_return}
   r_i(t) = \ln{P_i(t)} - \ln{P_i(t - \Delta t)}
\end{equation}
where $P_i(t)$ denotes the daily closing price of the agricultural future $i=1, \cdots, N$ at time $t=1, \cdots, T$, and the time scale $\Delta t$ equals 1 day.

The contract units of some agricultural futures have changed since they were listed, such as the Mustard Seed Futures and Castor Seed Futures of NCDEX and DCE's Fiberboard Futures. In order to ensure the objectivity and accuracy of the return series, we exclude the return data of all agricultural futures on these dates with contract units change.

Table~\ref{Tab:AgroFutures:Stat} presents the descriptive statistics of the return series for all agricultural futures. Generally, if the maximum return (Max) is large, the absolute minimum return ($-$Min) is also large. Most of the mean returns (Mean) are positive and the range of standard deviation (SD) is $[0.002,0.025]$. From Table~\ref{Tab:AgroFutures:Stat}, we find that the distributions of all returns are right- or left-skewed. Except for the Red Hard Winter Wheat Futures of CBOT, the kurtosis of other agricultural futures return series is significantly larger than 3, which is the kurtosis of Gaussian distributions, implying that the empirical return distributions have fat tails.

\begin{landscape}
\renewcommand\arraystretch{1.25}
\begin{longtable}{cccccrrrrrr}%[!t]
%   \centering
  \caption{Descriptive statistics of agricultural futures return series}\\
  \endfirsthead
  \caption{Descriptive statistics of agricultural futures return series (continued)}\\
  \hline 
% 符号表示 & 符号含义\\
% \cline{1-2}
\endhead
% \hline
    %   \small{
    % \begin{tabular}{cccccrrrrrr}
    \toprule
    Label & Futures & Exchange & Country & Listing Date & \multicolumn{1}{c}{Max} & \multicolumn{1}{c}{Min} & \multicolumn{1}{c}{Mean($\times 10^{4}$)} & \multicolumn{1}{c}{SD} & \multicolumn{1}{c}{Skewness} & \multicolumn{1}{c}{Kurtosis} \\
    \midrule
    1 & Corn & BMF & Brazil & 2008-11-18 & 0.273 & $-$0.263 & 2.435 & 0.013 & $-$0.390 & 104.685 \\
    2 & Soybean & BMF & Brazil & 2012-06-11 & 0.094 & $-$0.180 & $-$0.151 & 0.009 & $-$3.224 & 75.639 \\
    3 & Coffee Arabica & BMF & Brazil & 2000-01-03 & 0.234 & $-$0.285 & $-$0.110 & 0.020 & $-$0.017 & 13.660 \\
    4 & Soybean Oil & CBOT & USA & 2000-01-01 & 0.087 & $-$0.160 & 1.731 & 0.015 & $-$0.093 & 4.855 \\
    5 & Soybean Meal & CBOT & USA & 2000-01-01 & 0.369 & $-$0.330 & 1.995 & 0.019 & $-$0.900 & 56.097 \\
    6 & Corn & CBOT & USA & 2000-01-01 & 0.193 & $-$0.208 & 1.530 & 0.018 & 0.189 & 15.644 \\
    7 & Oats & CBOT & USA & 2002-09-17 & 0.192 & $-$0.189 & 1.804 & 0.021 & $-$0.229 & 13.934 \\
    8 & Rough Rice & CBOT & USA & 2003-11-19 & 0.099 & $-$0.463 & 0.925 & 0.015 & $-$5.478 & 179.657 \\
    9 & Soybean & CBOT & USA & 2000-01-04 & 0.089 & $-$0.449 & 0.135 & 0.016 & $-$4.623 & 122.176 \\
    10 & Wheat & CBOT & USA & 2000-01-01 & 0.139 & $-$0.151 & 1.693 & 0.019 & 0.178 & 4.216 \\
    11 & Red Hard Winter Wheat & CBOT & USA & 2000-01-03 & 0.088 & $-$0.111 & 1.410 & 0.018 & 0.040 & 2.077 \\
    12 & Cash Settled Butter & CME & USA & 2005-09-19 & 0.301 & $-$0.288 & $-$0.248 & 0.013 & $-$1.122 & 155.363 \\
    13 & Class \Rmnum{3} Milk & CME & USA & 2000-01-03 & 0.489 & $-$0.427 & 0.904 & 0.020 & $-$0.334 & 170.846 \\
    14 & Cash Settled Cheese & CME & USA & 2010-06-21 & 0.467 & $-$0.434 & 0.217 & 0.014 & 0.284 & 460.659 \\
    15 & Non Fat Dry Milk & CME & USA & 2001-07-06 & 0.156 & $-$0.233 & 0.211 & 0.011 & 1.064 & 95.621 \\
    16 & Live Cattle & CME & USA & 2010-12-07 & 0.126 & $-$0.085 & 0.197 & 0.008 & $-$0.065 & 27.449 \\
    17 & Feeder Cattle & CME & USA & 2000-01-03 & 0.094 & $-$0.086 & 0.903 & 0.010 & $-$0.317 & 10.847 \\
    18 & Lean Hogs & CME & USA & 2000-01-03 & 0.236 & $-$0.272 & 0.363 & 0.023 & $-$0.736 & 26.065 \\
    19 & Wheat WH & CZCE & China & 2003-03-28 & 0.142 & $-$0.146 & 0.751 & 0.009 & 0.545 & 59.456 \\
    20 & Cotton No.1 & CZCE & China & 2004-06-01 & 0.092 & $-$0.160 & $-$0.008 & 0.010 & $-$0.709 & 19.080 \\
    21 & White Sugar & CZCE & China & 2006-01-06 & 0.097 & $-$0.078 & 0.184 & 0.010 & 0.502 & 9.117 \\
    22 & Rapeseed Oil & CZCE & China & 2007-06-08 & 0.078 & $-$0.080 & 0.026 & 0.010 & $-$0.723 & 11.421 \\
    23 & Early Rice & CZCE & China & 2009-04-20 & 0.095 & $-$0.179 & 0.487 & 0.007 & $-$1.337 & 93.582 \\
    24 & Wheat PM & CZCE & China & 2012-01-17 & 0.165 & $-$0.094 & 0.112 & 0.008 & 1.538 & 64.689 \\
    25 & Rapeseed Meal & CZCE & China & 2012-12-28 & 0.072 & $-$0.144 & 0.337 & 0.009 & $-$1.275 & 29.078 \\
    26 & Rapeseed & CZCE & China & 2012-12-28 & 0.152 & $-$0.132 & 0.030 & 0.009 & 0.672 & 80.410 \\
    27 & Japonica Rice & CZCE & China & 2013-11-18 & 0.115 & $-$0.110 & $-$0.407 & 0.007 & 0.567 & 68.366 \\
    28 & Late Indica Rice & CZCE & China & 2014-07-08 & 0.120 & $-$0.121 & 0.124 & 0.007 & $-$0.334 & 99.539 \\
    29 & Cotton Yarn & CZCE & China & 2017-08-18 & 0.080 & $-$0.063 & $-$0.104 & 0.004 & 0.145 & 69.451 \\
    30 & Chinese Jujube & CZCE & China & 2019-04-30 & 0.085 & $-$0.052 & 0.224 & 0.004 & 3.286 & 129.710 \\
    31 & Apple & CZCE & China & 2017-12-22 & 0.205 & $-$0.394 & $-$0.363 & 0.009 & $-$14.090 & 848.168 \\
    32 & No.1 Soybean & DCE & China & 2000-01-04 & 0.108 & $-$0.124 & 1.975 & 0.012 & $-$0.183 & 9.517 \\
    33 & Soybean Meal & DCE & China & 2000-07-17 & 0.110 & $-$0.137 & 0.974 & 0.014 & $-$0.661 & 10.680 \\
    34 & Corn & DCE & China & 2004-09-22 & 0.123 & $-$0.157 & 1.502 & 0.008 & $-$1.811 & 77.873 \\
    35 & No.2 Soybean & DCE & China & 2004-12-22 & 0.201 & $-$0.179 & 0.871 & 0.013 & $-$0.340 & 42.301 \\
    36 & Soybean Oil & DCE & China & 2006-01-09 & 0.102 & $-$0.100 & 0.806 & 0.010 & $-$0.402 & 8.626 \\
    37 & RBD Palm Olein & DCE & China & 2007-10-29 & 0.095 & $-$0.089 & $-$0.349 & 0.011 & $-$0.246 & 6.697 \\
    38 & Egg & DCE & China & 2013-11-08 & 0.299 & $-$0.197 & 0.090 & 0.012 & 5.412 & 182.692 \\
    39 & Corn Starch & DCE & China & 2014-12-19 & 0.093 & $-$0.135 & 0.172 & 0.006 & $-$2.699 & 112.668 \\
    40 & Blockboard & DCE & China & 2013-12-06 & 0.413 & $-$0.388 & 1.160 & 0.021 & $-$2.110 & 144.496 \\
    41 & Fiberboard & DCE & China & 2013-12-06 & 0.457 & $-$0.535 & 0.256 & 0.021 & $-$2.915 & 239.098 \\
    42 & Polished Round$-$grained Rice & DCE & China & 2019-08-16 & 0.045 & $-$0.022 & $-$0.037 & 0.002 & 8.600 & 240.974 \\
    43 & Corn & EUREX & Europe & 2000-01-03 & 0.153 & $-$0.356 & 0.909 & 0.013 & $-$3.524 & 113.164 \\
    44 & Milling Wheat & EUREX & Europe & 2000-01-03 & 0.131 & $-$0.240 & 0.923 & 0.013 & $-$1.131 & 35.369 \\
    45 & Rapeseed & EUREX & Europe & 2000-01-03 & 0.067 & $-$0.166 & 1.567 & 0.010 & $-$2.119 & 28.560 \\
    46 & Cocoa & ICE & USA & 2000-01-04 & 0.093 & $-$0.171 & 2.007 & 0.017 & $-$0.111 & 4.641 \\
    47 & Cotton No.2 & ICE & USA & 2000-01-01 & 0.136 & $-$0.153 & 0.743 & 0.017 & 0.144 & 4.729 \\
    48 & Feed Wheat & ICE & USA & 2000-01-04 & 0.165 & $-$0.172 & 1.850 & 0.012 & 0.390 & 23.432 \\
    49 & Coffee NY & ICE & USA & 2000-01-03 & 0.166 & $-$0.128 & 0.120 & 0.021 & 0.212 & 3.298 \\
    50 & Robusta Coffee & ICE & USA & 2000-01-04 & 0.129 & $-$0.474 & $-$0.145 & 0.020 & $-$3.320 & 75.038 \\
    51 & White Sugar & ICE & USA & 2000-01-04 & 0.140 & $-$0.155 & 1.634 & 0.017 & $-$0.705 & 7.678 \\
    52 & Sugar No.11 & ICE & USA & 2000-01-01 & 0.172 & $-$0.195 & 1.691 & 0.021 & 0.084 & 6.126 \\
    53 & Wheat & ICE & USA & 2000-01-04 & 0.154 & $-$0.177 & 1.854 & 0.013 & 0.250 & 18.851 \\
    54 & Crude Palm Oil & MDEX & Malaysia & 2017-01-04 & 0.064 & $-$0.105 & 0.235 & 0.007 & $-$0.783 & 33.515 \\
    55 & Cotton & MCX & India & 2011-10-03 & 0.081 & $-$0.127 & 0.145 & 0.007 & $-$1.179 & 36.698 \\
    56 & Kapas & MCX & India & 2004-01-22 & 0.243 & $-$0.232 & 1.564 & 0.012 & 4.304 & 119.357 \\
    57 & Cardamom & MCX & India & 2006-02-14 & 0.310 & $-$0.308 & 3.155 & 0.025 & 1.010 & 25.021 \\
    58 & Castor Seeds & MCX & India & 2004-01-22 & 0.312 & $-$0.377 & 1.629 & 0.010 & $-$2.448 & 500.317 \\
    59 & Soybean & NCDEX & India & 2003-12-15 & 0.225 & $-$0.279 & 1.818 & 0.015 & $-$1.874 & 59.381 \\
    60 & Soy Oil & NCDEX & India & 2004-02-23 & 0.087 & $-$0.269 & 1.574 & 0.009 & $-$4.441 & 130.764 \\
    61 & Mustard Seed & NCDEX & India & 2003-12-15 & 0.188 & $-$0.159 & 1.615 & 0.012 & $-$1.377 & 40.705 \\
    62 & CHANA & NCDEX & India & 2017-07-14 & 0.047 & $-$0.073 & $-$0.371 & 0.006 & $-$0.807 & 25.715 \\
    63 & Cottonseed Oilcake & NCDEX & India & 2005-04-05 & 0.230 & $-$0.542 & 2.354 & 0.019 & $-$8.218 & 202.436 \\
    64 & Guarseed & NCDEX & India & 2014-05-15 & 0.100 & $-$0.070 & $-$0.442 & 0.010 & 0.296 & 10.901 \\
    65 & Guar Gum Refined Splits & NCDEX & India & 2004-07-27 & 0.222 & $-$0.347 & 2.190 & 0.019 & $-$1.175 & 31.585 \\
    66 & Turmeric & NCDEX & India & 2004-11-22 & 0.262 & $-$0.363 & 1.649 & 0.020 & $-$1.699 & 51.529 \\
    67 & Castor Seed & NCDEX & India & 2004-07-23 & 0.253 & $-$0.148 & 1.585 & 0.013 & 0.949 & 36.964 \\
    68 & Jeera & NCDEX & India & 2005-06-21 & 0.160 & $-$0.238 & 0.604 & 0.015 & $-$0.061 & 19.886 \\
    69 & Coriander & NCDEX & India & 2008-08-11 & 0.147 & $-$0.382 & $-$1.076 & 0.017 & $-$2.543 & 66.795 \\
    70 & Shankar Kapas & NCDEX & India & 2011-04-11 & 0.305 & $-$0.187 & $-$0.381 & 0.009 & 6.109 & 372.963 \\
    71 & Milk & NZX & New Zealand & 2016-09-09 & 0.075 & $-$0.113 & 0.519 & 0.003 & $-$6.455 & 735.723 \\
    72 & Skim Milk Powder & NZX & New Zealand & 2011-02-18 & 0.323 & $-$0.179 & $-$0.518 & 0.012 & 2.025 & 168.799 \\
    73 & Whole Milk Powder & NZX & New Zealand & 2010-10-08 & 0.325 & $-$0.219 & $-$0.151 & 0.013 & 2.849 & 141.566 \\
    74 & ASX Feed Barley & SFE & Australia & 2003-05-26 & 0.306 & $-$0.511 & 0.331 & 0.020 & $-$3.770 & 127.921 \\
    \bottomrule
    % \end{tabular}
    % }%

  \label{Tab:AgroFutures:Stat}%
\end{longtable}%
    \end{landscape}
% 2021/06/18, 12:22

\section{Empirical analysis}
\label{S1:EmpAnal}

\subsection{Correlation structure of the global agricultural futures market}

\subsubsection{Distribution of correlation coefficients} 

Since different agricultural futures have varying levels of volatility, we define a standardized return for simplicity
\begin{equation}\label{Eq:Standardized_return}
   g_i(t) = \frac{r_i(t) - \langle r_i \rangle}{\sigma_i}
\end{equation}
where $\langle \cdot \rangle$ denotes the mean value of a given time series and $\sigma_i = \sqrt{\langle r_i^2 \rangle - \langle r_i \rangle ^2}$ is the standard deviation of $r_i$. The correlation coefficient $c_{ij}$ between $g_i(t)$ and $g_j(t)$ can be computed as follows:
\begin{equation}\label{Eq:Correlation_coefficient}
   c_{ij} = \left\langle g_i(t)g_j(t) \right\rangle,
\end{equation}
which forms the correlation matrix $\mathbf{C}$.
By definition, the correlation coefficients $c_{ij}$ are restricted to the interval $[-1,1]$, where $c_{ij}=1$ corresponds to perfect positive correlation, $c_{ij}=-1$ corresponds to perfect negative correlation, and $c_{ij}=0$ implies that there is no correlation between $g_i(t)$ and $g_j(t)$.

\begin{figure}[!t]
\centering
\includegraphics[width=0.5\linewidth]{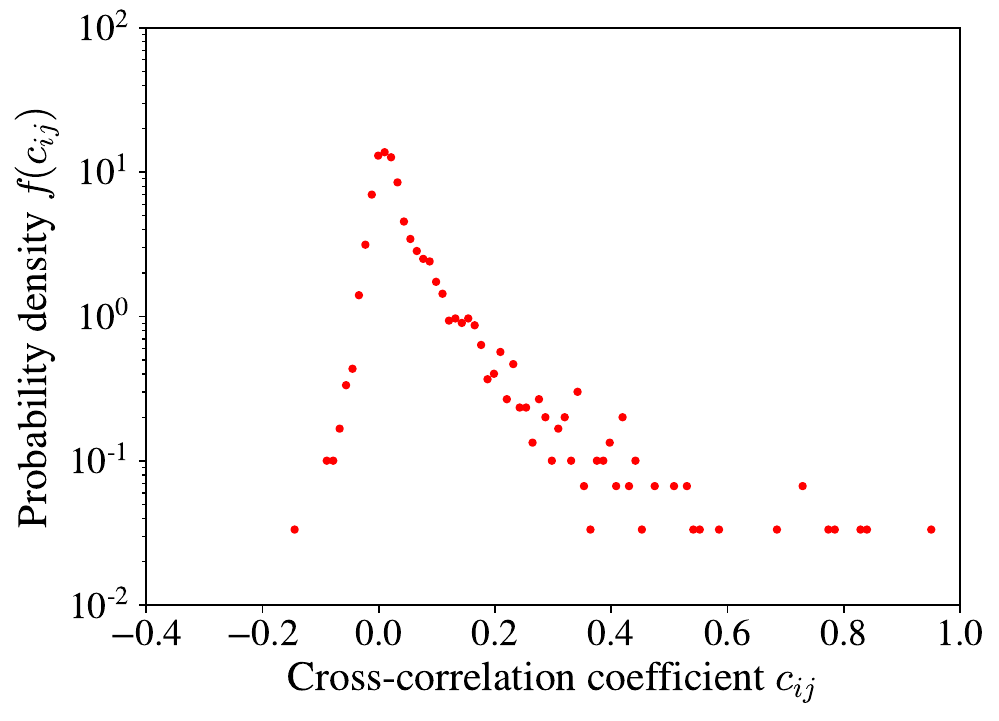}
\caption{Distribution of the correlation coefficients $c_{ij}$ of the return time series of agricultural futures.}
\label{Fig:AgriFut_PDF_corrCoefficient}
\end{figure}

Figure~\ref{Fig:AgriFut_PDF_corrCoefficient} shows the distribution $f(c_{ij})$ of correlation coefficients $c_{ij}$. We note that $f(c_{ij})$ is asymmetric and centered around a positive mean value, that is $\langle c_{ij} \rangle > 0$, indicating that positive correlations are more prevalent in the correlation structure of the agricultural futures market. There are also pairs of agricultural futures whose returns are highly correlated with the correlation coefficients close to 1. In addition, the probability distribution of correlation coefficients is leptokurtic and fat tailed.

\subsubsection{Distribution of eigenvalues} 

For the empirical correlation matrix $\mathbf{C}$ of agricultural futures returns, we can calculate its eigenvalues and their corresponding eigenvectors by solving the following equation
\begin{equation}\label{Eq:Eigenvalue}
   \mathbf{C} = \mathbf{U \Lambda U}^{\mathrm{T}},
\end{equation}
where $\mathbf{U}$ denotes matrix of the eigenvectors $\mathbf{u}_i$ and $\mathbf{\Lambda}$ denotes the diagonal matrix of the eigenvalues $\lambda_i$ and $\mathbf{U}^{\mathrm{T}}$ is the transpose of $\mathbf{U}$. The probability density function $f_\mathbf{C}(\lambda)$ of $\mathbf{\Lambda}$ is given by
\begin{equation}\label{Eq:Probability_density_C}
   f_\mathbf{C}(\lambda) = \frac{1}{N} \frac{\text{d}n(\lambda)}{\text{d}\lambda}
\end{equation}
where $n(\lambda)$ denotes the number of eigenvalues that are smaller than $\lambda$ of the matrix $\mathbf{C}$.

We consider a random matrix $\mathbf{R}$ given by
\begin{equation}\label{Eq:Random_matrix}
   \mathbf{R} = \frac{1}{T} \mathbf{A} \mathbf{A}^{\mathrm{T}}
\end{equation}
where $\mathbf{A}$ is an $N \times T$ matrix that contains $N$ time series of $T$ random elements with zero mean and unit variance, which are mutually uncorrelated. According to \cite{Plerou-Gopikrishnan-Rosenow-Amaral-Guhr-Stanley-2002-PhysRevE}, the statistical properties of the random matrix $\mathbf{R}$ are known. Particularly, in the limit $N \to \infty$, $T \to \infty$, such that $Q = T/N>1$ is fixed, the probability density function $f_{\text{RMT}}(\lambda)$ of eigenvalues $\lambda$ of $\mathbf{R}$ is given by
\begin{equation}\label{Eq:Probability_density_RMT}
   f_{\text{RMT}}(\lambda) = \frac{Q}{2\pi} \frac{\sqrt{\left(\lambda_{\max}^{\text{RMT}}-\lambda\right)\left(\lambda-\lambda_{\min}^{\text{RMT}}\right)}}{\lambda}
\end{equation}
with $\lambda \in \left[\lambda_{\min}^{\text{RMT}}, \lambda_{\max}^{\text{RMT}}\right]$, where $\lambda_{\min}^{\text{RMT}}$ and $\lambda_{\max}^{\text{RMT}}$ are the minimum and maximum eigenvalues of $\mathbf{R}$, respectively, given by
\begin{equation}\label{Eq:Minimum_eigenvalue_RMT}
   \lambda_{\min}^{\text{RMT}} = 1 + \frac{1}{Q} - 2\sqrt{\frac{1}{Q}}
\end{equation}
and
\begin{equation}\label{Eq:Maximum_eigenvalue_RMT}
   \lambda_{\max}^{\text{RMT}} = 1 + \frac{1}{Q} + 2\sqrt{\frac{1}{Q}}.
\end{equation}

In this paper, $N$ denotes the quantity of agricultural futures, whose value is 74, and $T$ denotes the length of return series, whose value is 5473. Thus $Q=5473/74=73.959$, and we obtain the
smallest eigenvalue $\lambda_{\min}^{\text{RMT}}=0.781$ and the largest eigenvalue $\lambda_{\max}^{\text{RMT}}=1.246$ from Eq. (\ref{Eq:Minimum_eigenvalue_RMT}) and Eq. (\ref{Eq:Maximum_eigenvalue_RMT}). We calculate the $N$ eigenvalues of the empirical correlation matrix $\mathbf{C}$, where $\lambda_{i}$ are sorted in descending order ($\lambda_{i}>\lambda_{i+1}$). In particular, the smallest eigenvalue $\lambda_{74}=0.043$, which is about 1/18 of $\lambda_{\min}^{\text{RMT}}=0.781$, and the largest eigenvalue $\lambda_{1}=6.465$, which is about 5 times larger than $\lambda_{\max}^{\text{RMT}}=1.246$.

\begin{figure}[!t]
\centering
 \includegraphics[width=0.5\linewidth]{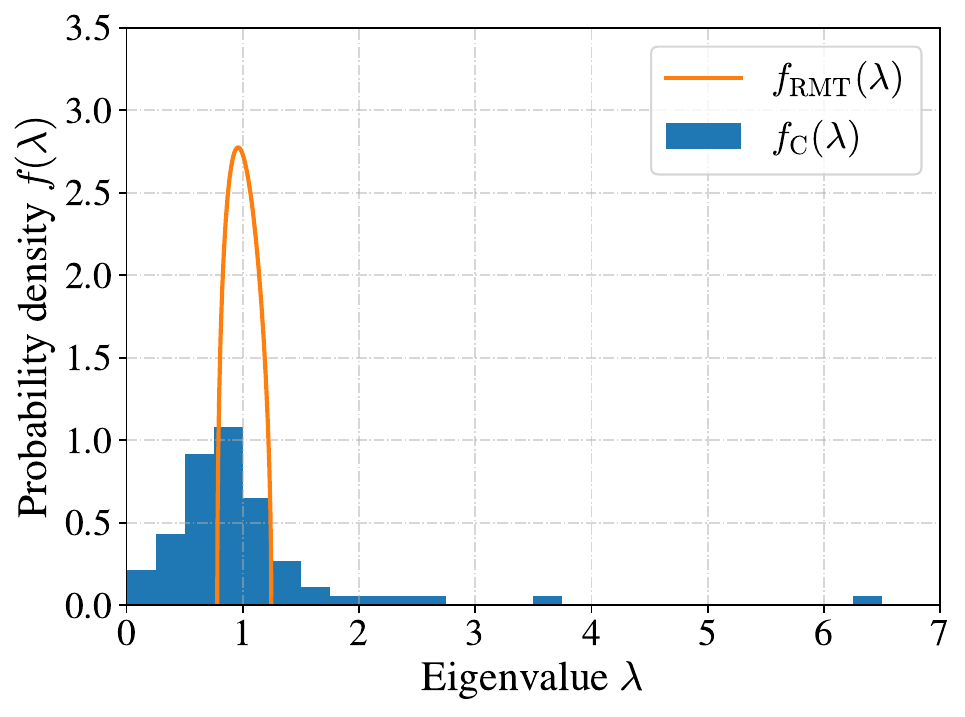}
\caption{Empirical distribution of eigenvalues of the correlation matrix $\mathbf{C}$. The solid curve is the theoretical distribution $f_{\text{RMT}}(\lambda)$ given by Eq.~(\ref{Eq:Probability_density_RMT}) of random matrix theory (RMT).}
\label{Fig:AgriFut_PDF_eigenvalue}
\end{figure}

Figure~\ref{Fig:AgriFut_PDF_eigenvalue} compares the empirical distribution $P_{\mathbf{C}}(\lambda)$ of these eigenvalues of $\mathbf{C}$ with the theoretical distribution $f_{\text{RMT}}(\lambda)$ on the basis of the random matrix $\mathbf{R}$. We note that a considerable proportion of the eigenvalues of $\mathbf{C}$ deviate from $f_{\text{RMT}}(\lambda)$. Specifically, 30 minimum eigenvalues ($\lambda_{45}$--$\lambda_{74}$) are smaller than $\lambda_{\min}^{\text{RMT}}$, and 13 maximum eigenvalues ($\lambda_{1}$--$\lambda_{13}$) are larger than $\lambda_{\max}^{\text{RMT}}$, while 31 eigenvalues ($\lambda_{14}$--$\lambda_{44}$) fall within the bulk of $f_{\text{RMT}}(\lambda)$.

% 2021/11/17, 10:00, the whole subsection revised

\subsection{Financial implications of eigenvectors}

\subsubsection{Inverse participation ratio} %逆参与率

In order to reflect the degree of deviation from RMT results for the correlation matrix $\mathbf{C}$, we investigate the inverse participation ratio (IPR). The inverse participation ratio quantifies the reciprocal of the number of eigenvector components that contribute significantly. Thus, the larger the value of IPR, the less the number of contributing components. For the $k$-th eigenvector $\mathbf{u}^k$, its IPR can be calculated as follows
\begin{equation}\label{Eq:IPR}
   I^k = \sum\limits_{l=1}^{N}\left[u_l^k\right]^4,
\end{equation}
where $\left\{u_l^k: l,k=1, \cdots, N\right\}$ are the components of the $k$-th eigenvector $\mathbf{u}^k$. If all the components of $\mathbf{u}^k$ are the same, that is, $u_l^k=1/\sqrt{N}$, then $I^k=1/N$ so that the number of contributing components is $N$, meaning that all the components contribute equally. If $\mathbf{u}^k$ has only one non-zero component, then $I^k=1$.

\begin{figure}[!ht]
\centering
\includegraphics[width=0.5\linewidth]{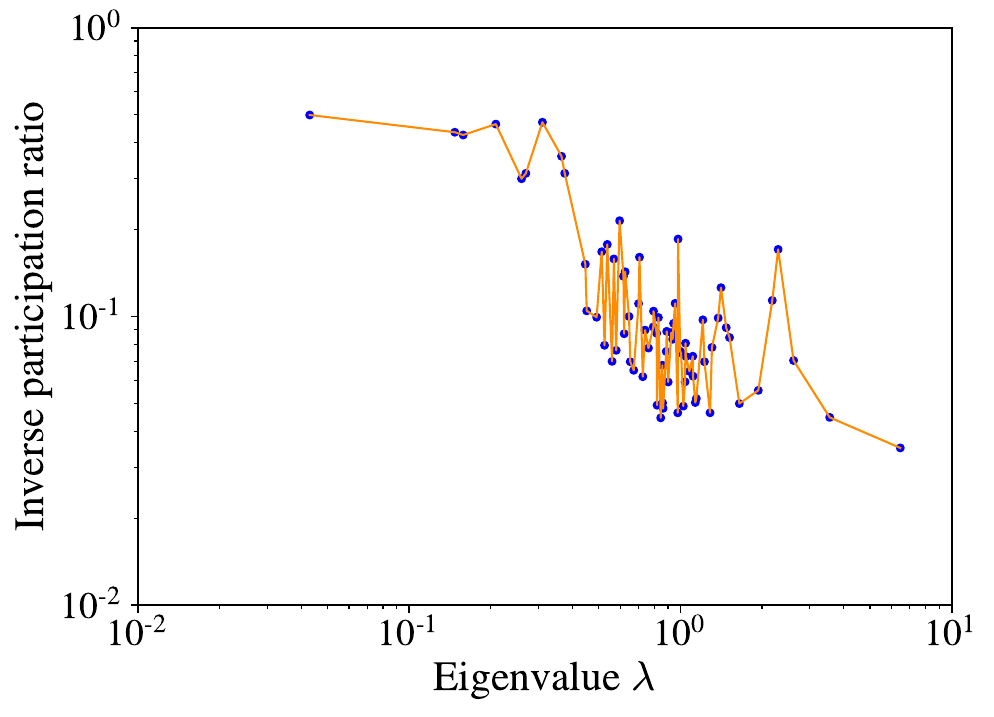}
\caption{Inverse participation ratios of the eigenvalues for the correlation matrix $\mathbf{C}$.}
\label{Fig:AgriFut_IPR}
\end{figure}

Figure~\ref{Fig:AgriFut_IPR} shows the IPRs corresponding to all eigenvalues of the correlation matrix $\mathbf{C}$. The average value $\langle I \rangle$ is 0.127, indicating that the average number of contributing components is about 9. The IPRs of the eigenvalues deviating from the RMT prediction exhibit significant deviations from $\langle I \rangle$. The largest eigenvalue $\lambda_1$ has ${I^1}=0.035$, and its reciprocal is about 29, showing that about 29 agricultural futures participate in its corresponding eigenvector $\mathbf{u}^1$. For other largest eigenvalues which fall outside the RMT upper bound $\lambda_{\max}^{\text{RMT}}$, their $I^k$ values are quite different, suggesting that varying numbers of agricultural futures make important contributions to their corresponding eigenvectors.

We also notice that some small eigenvalues less than the RMT lower bound $\lambda_{\min}^{\text{RMT}}$ have large $I^k$ values, which indicates that only a few agricultural futures contribute to them. For the smallest deviating eigenvalues, the values of $I$ is about 0.5 so that there are only 2 return time series involved. In other words, the behavior of these smallest eigenvalues is determined by asset pairs, as verified by the U.S. stock market \citep{Plerou-Gopikrishnan-Rosenow-Amaral-Guhr-Stanley-2002-PhysRevE} and the global oil spot market \citep{Dai-Xie-Jiang-Jiang-Zhou-2016-EmpirEcon}.

% 2021/11/17, 10:37

\subsubsection{Eigenportfolios and the collective market effect}

Based on the above analysis, it can be found that there are significant differences between the statistical properties of the correlation matrix $\mathbf{C}$ and the random matrix $\mathbf{R}$, which implies that $\mathbf{C}$ is not a random correlation matrix, and its deviating eigenvalues contain economic information.

To reveal the information contained in these deviating eigenvalues of $\mathbf{C}$, we propose to construct eigenportfolios for each eigenvalue. For $\lambda_k$ , the return of its corresponding eigenportfolio at time $t$ is given by
\begin{equation}\label{Eq:Eigenportfolio_return}
   G^k(t) = {\sum\limits_{j=1}^{N}u_{j}^{k}r_j(t)} \Bigg/ {\sum\limits_{j=1}^{N}u_{j}^{k}}
\end{equation}
where $\sum\limits_{j=1}^{N}u_{j}^{k}r_j(t)$ is the projection of the agricultural futures return series $r_j(t)$ on the $k$-th eigenvector $\mathbf{u}^k$.

\begin{figure}[!t]
\centering
\includegraphics[width=0.321\linewidth]{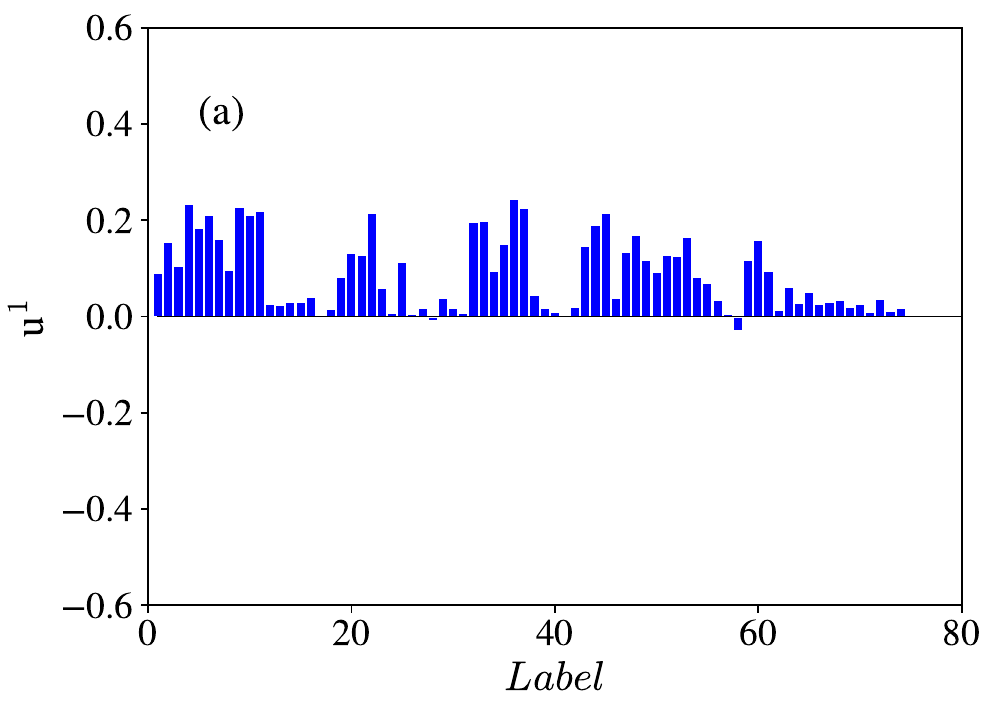}
\includegraphics[width=0.321\linewidth]{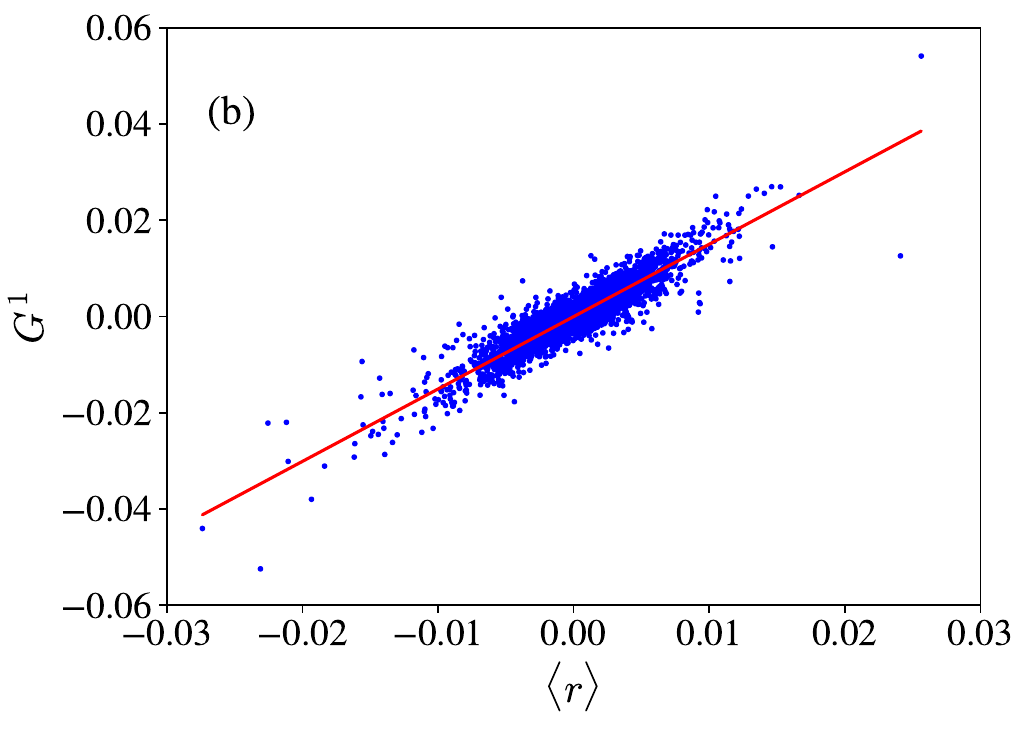}
\includegraphics[width=0.321\linewidth]{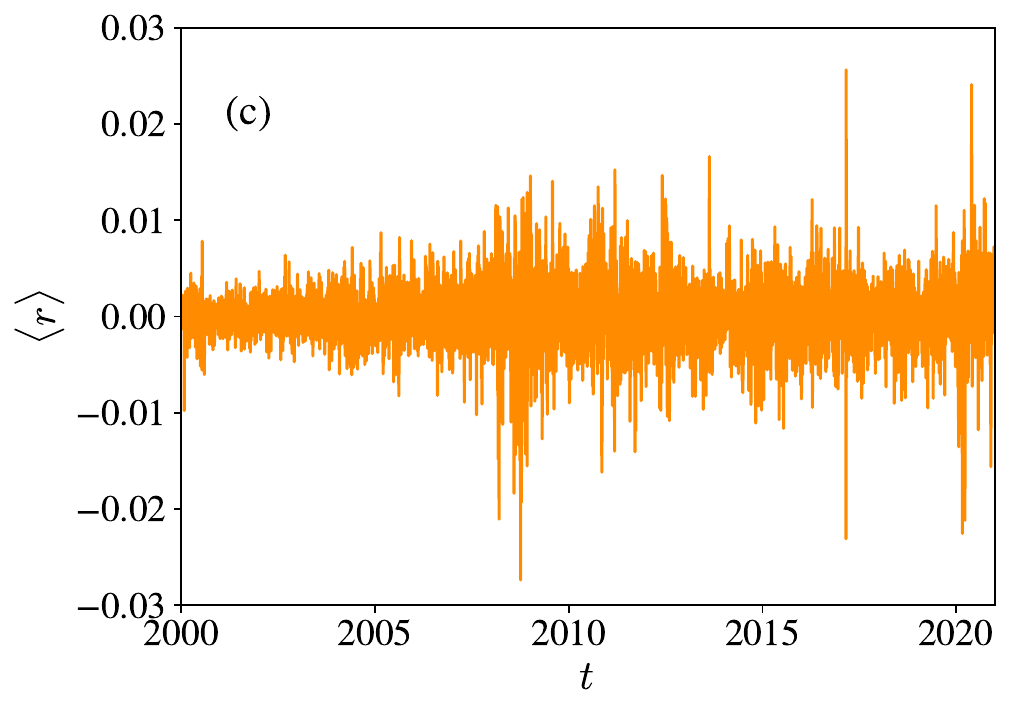}
\includegraphics[width=0.321\linewidth]{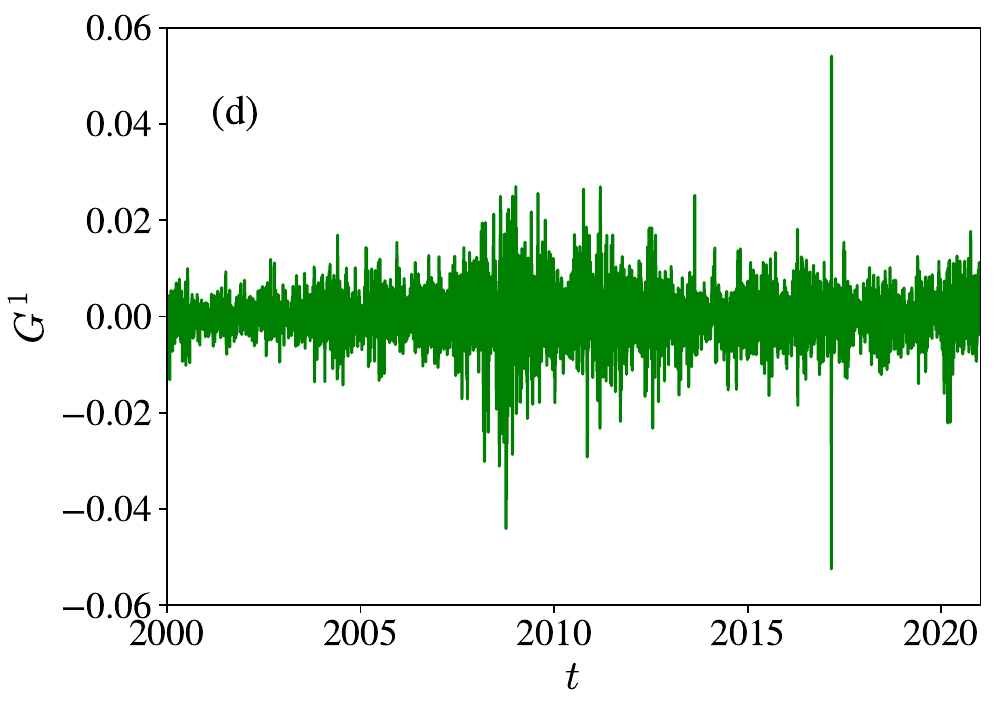}
\includegraphics[width=0.321\linewidth]{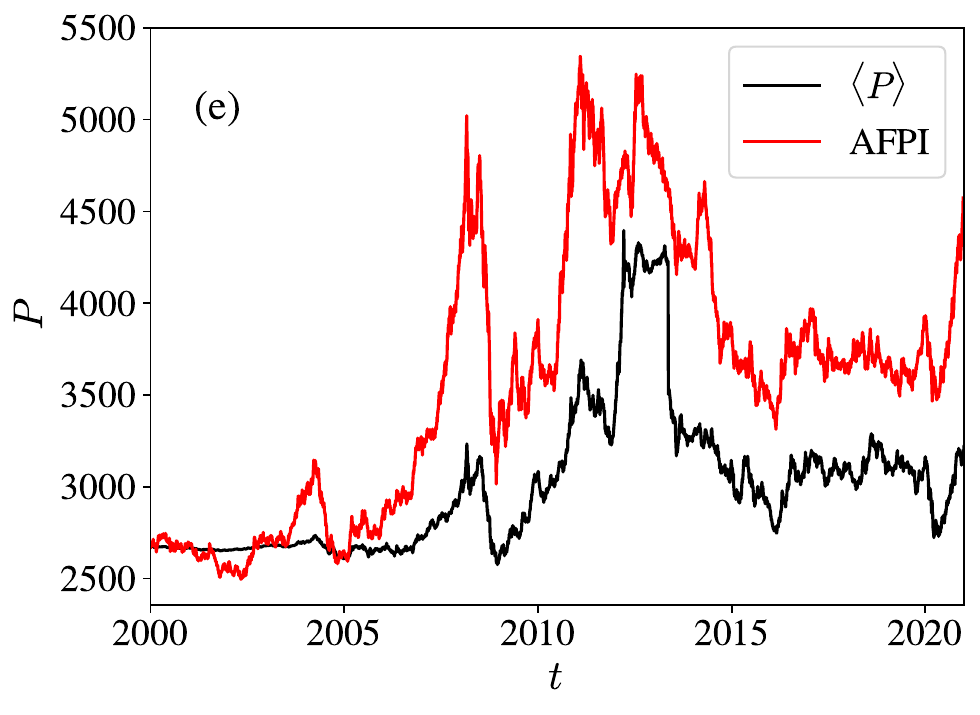}
\includegraphics[width=0.321\linewidth]{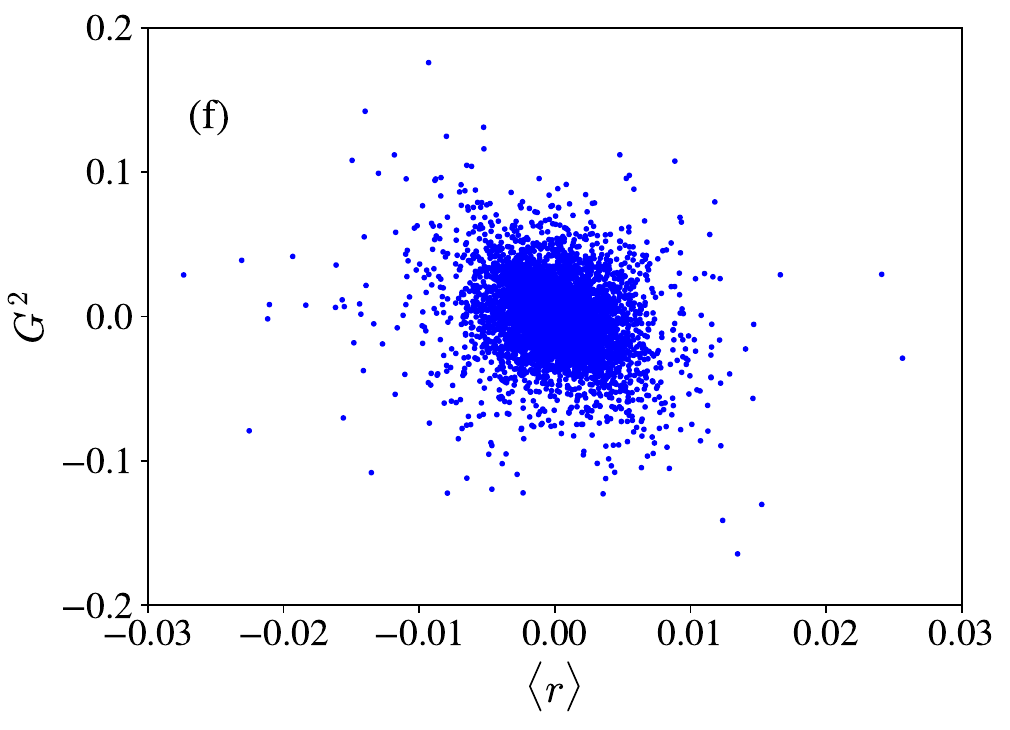}
\includegraphics[width=0.321\linewidth]{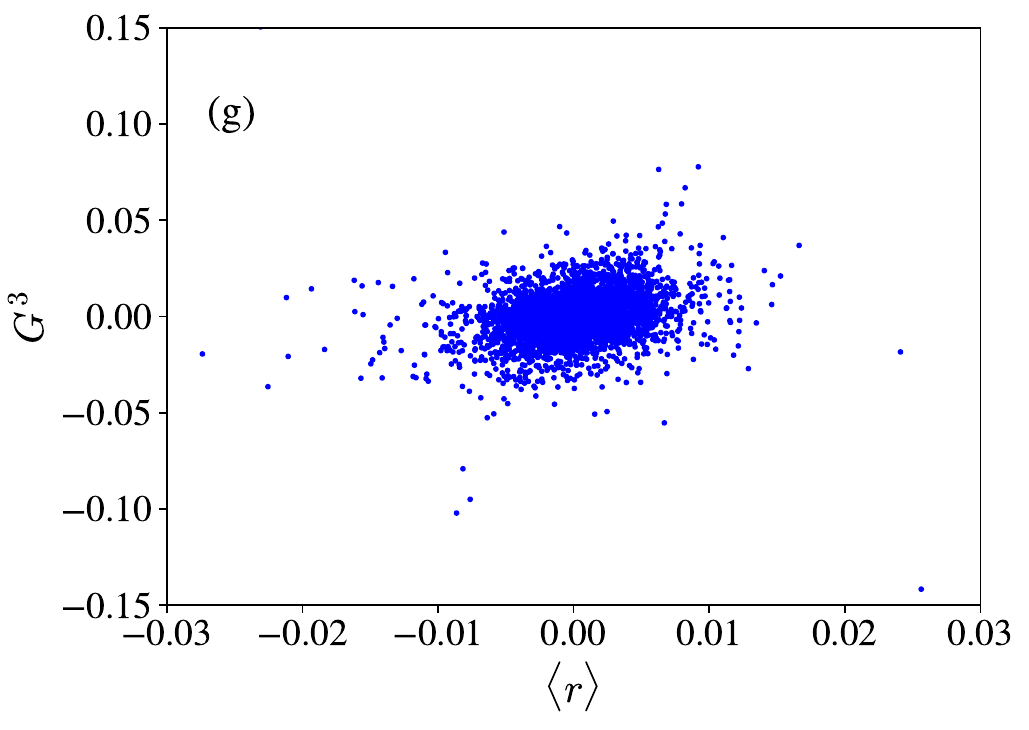}
\includegraphics[width=0.321\linewidth]{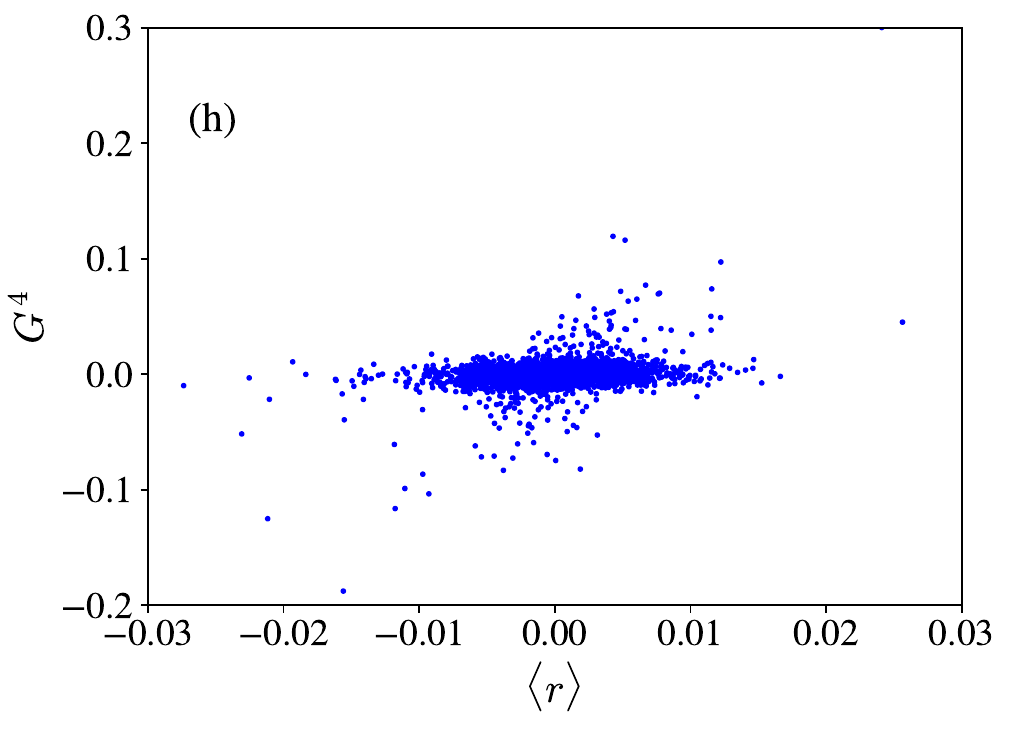}
\includegraphics[width=0.321\linewidth]{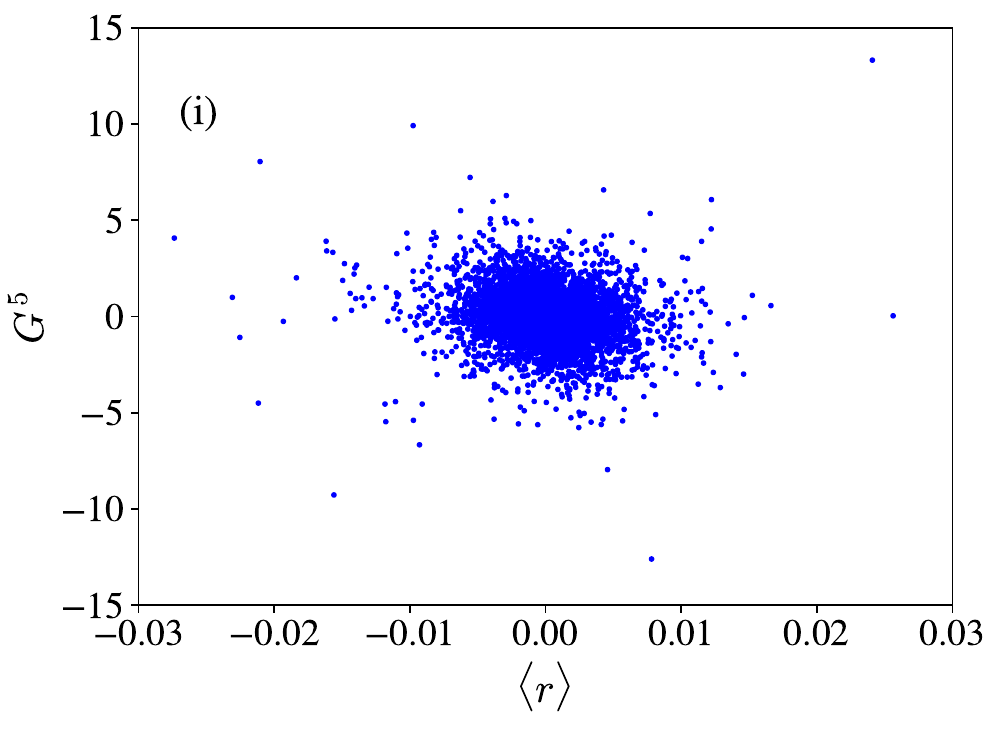}
\caption{Testing the collective market effect of the largest eigenvalue. (a) Components of the eigenvector $\mathbf{u}^1$ of the largest eigenvalue $\lambda_1$. (b) Relationship between the eigenportfolio returns $G^{1}$ and the mean returns $\langle{r}\rangle$. (c) Evolution of the mean returns $\langle{r}\rangle$. (d) Evolution of the eigenportfolio returns $G^{1}$. (e) Evolution of the average daily price $\langle{P}\rangle$ and the constructed index AFPI based on the eigenportfolio associated with the largest eigenvalue. (f) Relationship between the eigenportfolio returns $G^{2}$ for $\lambda_2$ and the mean returns $\langle{r}\rangle$. (g) Relationship between the eigenportfolio returns $G^{3}$ for $\lambda_3$ and the mean returns $\langle{r}\rangle$. (h) Relationship between the eigenportfolio returns $G^{4}$ for $\lambda_4$ and the mean returns $\langle{r}\rangle$. (i) Relationship between the eigenportfolio returns $G^{5}$ for $\lambda_5$ and the mean returns $\langle{r}\rangle$.}
\label{Fig:AgriFut_u1component}
\end{figure}

Figure~\ref{Fig:AgriFut_u1component}(a) shows that most of the components of $\mathbf{u}^1$ have the same sign, that is positive. In addition, Fig.~\ref{Fig:AgriFut_u1component}(b) exhibits the relationship between the eigenportfolio returns $G^1$ and the mean returns $\langle r \rangle$ of all the 74 agricultural futures. There is an excellent linear relationship between $G^1$ and $\langle{r}\rangle$ with a quite high $R^2$ of 0.846. Therefore, the largest eigenvalue $\lambda_1$ of the correlation matrix $\mathbf{C}$ can be considered to reflect a collective market effect of the global agricultural futures market.

We further explore the evolution of the eigenportfolio returns $G^1$ based on the largest eigenvalue $\lambda_1$. The time series of $\langle r \rangle$ and $G^1$ from 2000 to 2020 are respectively illustrated in Fig.~\ref{Fig:AgriFut_u1component}(c) and Fig.~\ref{Fig:AgriFut_u1component}(d), and their trends seem to be quite similar. We also find that the fluctuations of $\langle r \rangle$ and $G^1$ are relatively large during 2008-2012 and 2017-2018, which indicates that the 2008 food crisis, the financial crisis, the China-US trade war and the US-Canada trade war have caused a huge impact on the global agricultural futures market. In addition, the fluctuations of agricultural futures returns have also increased since 2020, reflecting the instability of the international agricultural futures market to a certain extent and probably the impact of the COVID-19 pandemic.

Considering that the largest eigenvalue reflects the market effect common to all agricultural futures, we construct an agricultural futures price index (AFPI) for the global agricultural futures market on the basis of the eigenportfolio corresponding to $\lambda_1$ :
\begin{equation}\label{Eq:AFPI}
   \text{AFPI}(t) = \langle P(0) \rangle \exp \left[ \sum\limits_{t=1}^{T}G^1(t) \right]
\end{equation}
where $T$ is the length of the price series, and $\langle P(0) \rangle=2673.995$ is the average price on January 3, 2000. Figure~\ref{Fig:AgriFut_u1component}(e) compares the constructed index AFPI with the average price $\langle P \rangle$. It can be found that the evolution of AFPI has similar trends as $\langle P \rangle$. Moreover, AFPI is significantly greater than $\langle P \rangle$ for most of the time, showing that the index AFPI always performs better than $\langle P \rangle$. In other words, the eigenportfolio based on the largest eigenvalue outperforms the equally weighted portfolio under
the buy-and-hold strategy. Of particular interest is the steeper rise of AFPI since 2005, giving a more evident signal for the food crisis in 2008.

We further test whether other large deviating eigenvalues also have market effects. Plots (f--i) of Fig.~\ref{Fig:AgriFut_u1component} illustrate the relationship between the eigenportfolio returns $G^2$ for $\lambda_2$, $G^3$ for $\lambda_3$, $G^4$ for $\lambda_4$ and $G^5$ for $\lambda_5$ against the mean returns $\langle{r}\rangle$ respectively. Obviously, no evident linear relationship between the eigenportfolio returns and the mean returns can be observed for other large eigenvalues, which implies that these deviating eigenvalues do not bear any market-wide effects.

% 2021/11/17, 15:26

\subsubsection{Other largest eigenvalues and the sector-like effect} 

In order to avoid the impact of $\lambda_1$ on the returns of each agricultural futures $r_i(t)$, we remove the collective market effect embedded in the largest eigenvalue $\lambda_1$ according to the following steps. First of all, we perform the following linear regression and obtain the residuals $\epsilon_i(t)$:
\begin{equation}\label{Eq:Linear_regression}
   r_i(t) = \alpha_i+\beta_{i}M(t)+\epsilon_i(t),
\end{equation}
where $\langle \epsilon(t) \rangle=0$, $\langle M(t)\epsilon(t) \rangle=0$, $\alpha_i$ and $\beta_i$ are specific constants of the $i$-th agricultural futures, and $M(t)$ is an additive term common to all agricultural futures, which results in artificial correlations between any pair of agricultural futures. The decomposition of Eq. (\ref{Eq:Linear_regression}) lays a preliminary foundation for some pricing models that are widely used in economics, such as the capital asset pricing model and multi-factor models. Since $\mathbf{u}^1$ reflects an influence which is common to all agricultural futures, we can estimate the term $M(t)$ with $G^1(t)$.  Next, we calculate the correlation matrix $\mathbf{C}$ using $\epsilon_i(t)$ in Eq. (\ref{Eq:Standardized_return}) and Eq. (\ref{Eq:Correlation_coefficient}), then compute its eigenvalues and eigenvectors. Finally, we analyze other largest eigenvalues and their corresponding eigenvectors except the maximum eigenvalue.

\begin{figure}[!t]
\centering
\includegraphics[width=0.5\linewidth]{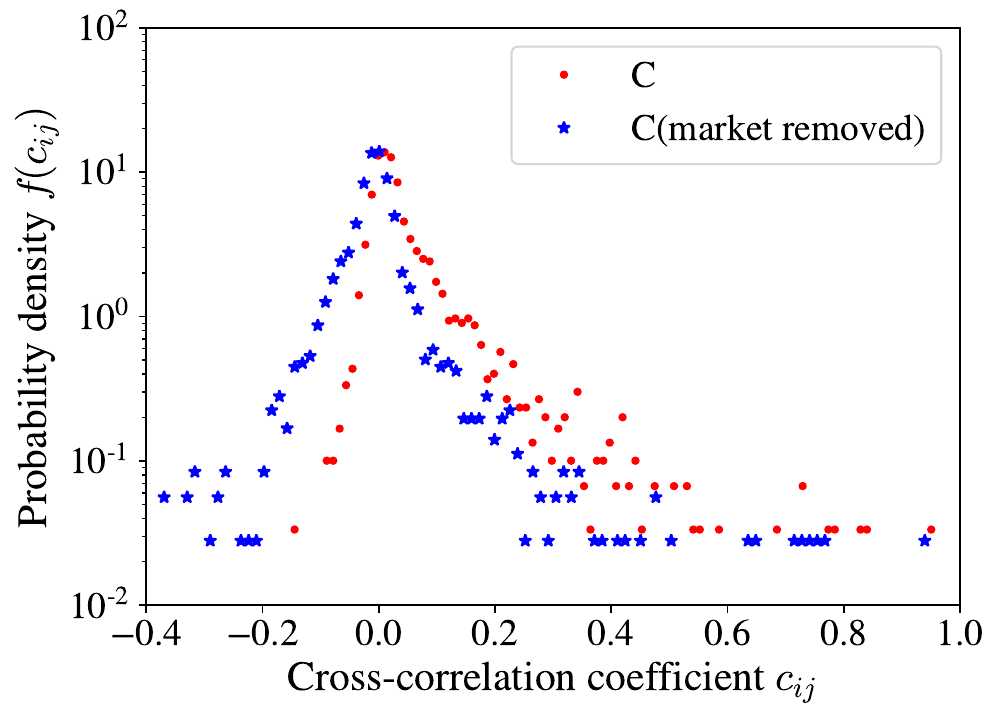}
\caption{Probability distribution $f(c_{ij})$ of the cross-correlation coefficients before and after removing the effect of the largest eigenvalue $\lambda_1$.}
\label{Fig:AgriFut_PDF_comparedCorrCoefficient}
\end{figure}

Figure~\ref{Fig:AgriFut_PDF_comparedCorrCoefficient} compares the probability distribution $f(c_{ij})$ of the correlation coefficients before and after removing the effect of the largest eigenvalue $\lambda_1$. We find that the latter is skewed relatively to the left, and it has a significantly smaller average value $\langle c_{ij} \rangle$, indicating that removing the effect of $\lambda_1$ shifts the correlation coefficients of agricultural futures toward smaller values generally. Therefore, a considerable degree of correlations embedded in $\mathbf{C}$ can be attributed to the effect of the maximum eigenvalue $\lambda_1$ and its associated eigenvector $\mathbf{u}^1$, which further supports the market effect of the maximum eigenvalue.

\begin{figure}[!t]
\centering
\includegraphics[width=0.4\linewidth]{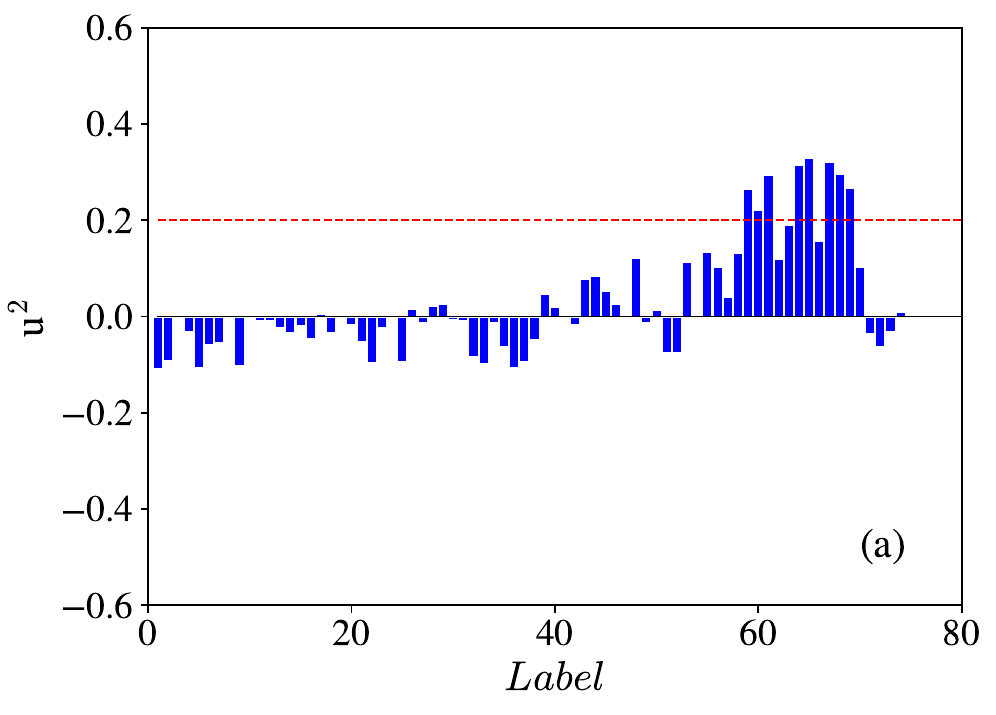}
\includegraphics[width=0.4\linewidth]{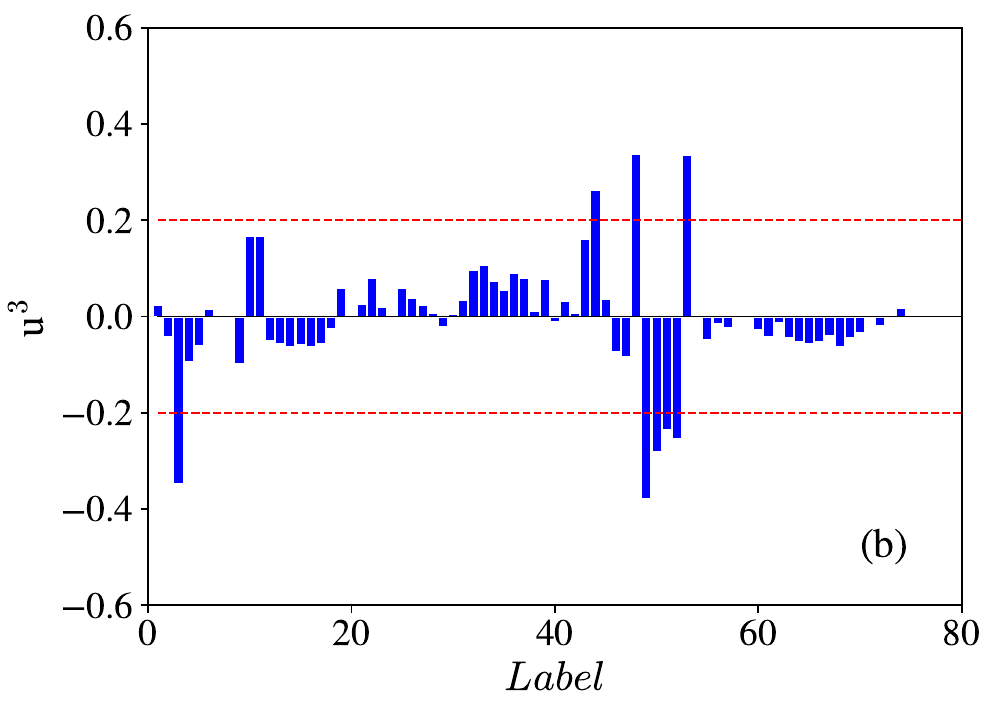}
\includegraphics[width=0.4\linewidth]{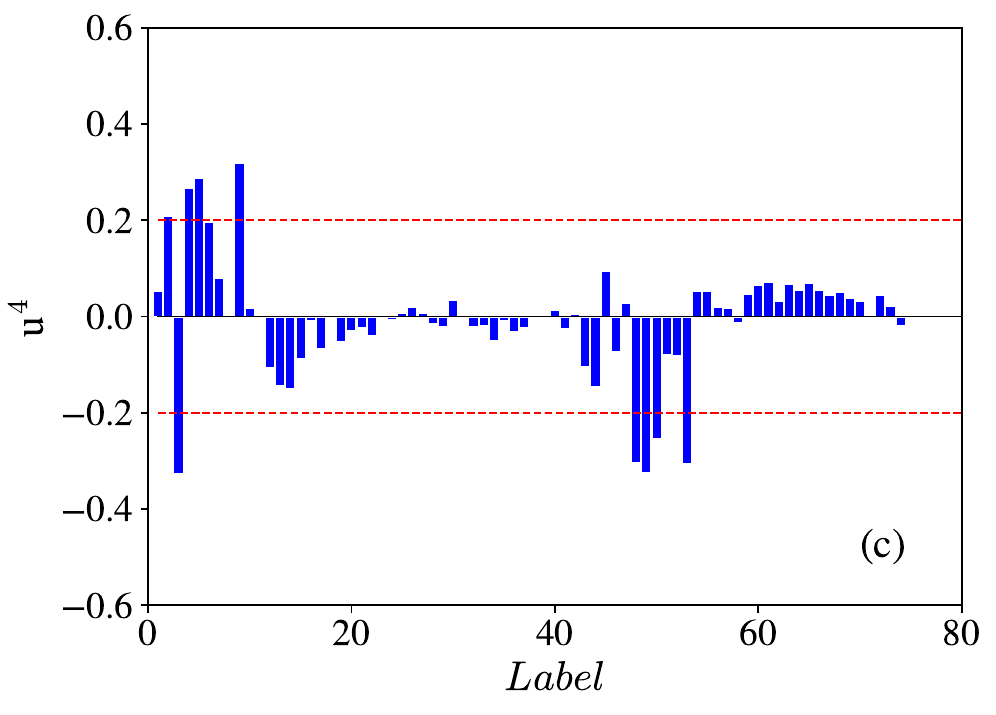}
\includegraphics[width=0.4\linewidth]{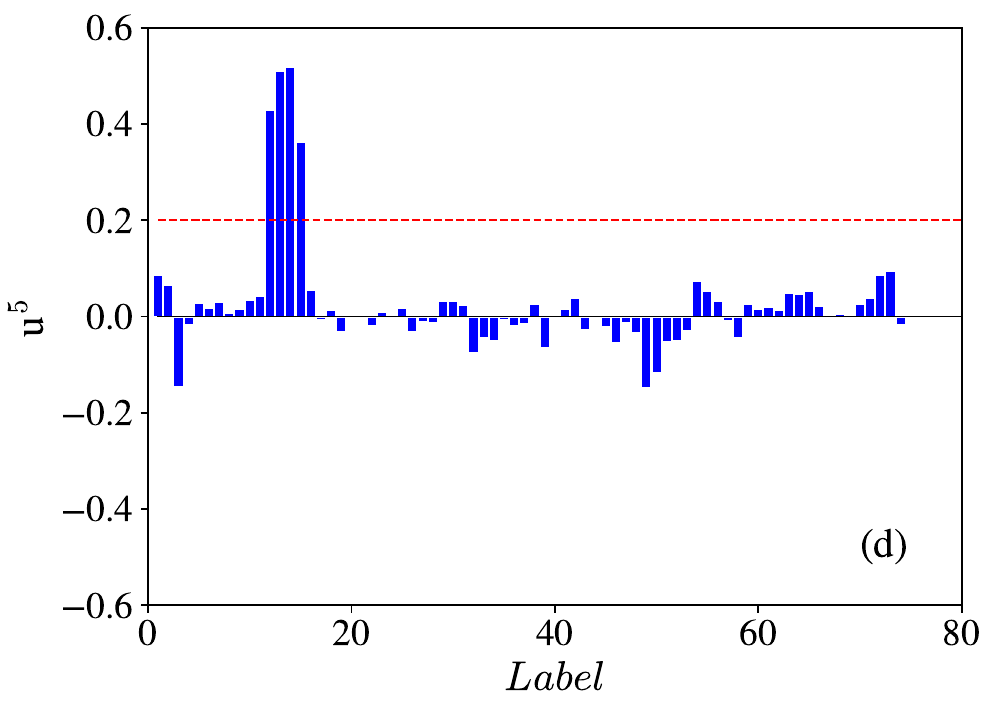}
\caption{Sector-like effect of the eigenvectors $\mathbf{u}^2$ (a), $\mathbf{u}^3$ (b), $\mathbf{u}^4$ (c) and $\mathbf{u}^5$ (d) corresponding to the four largest deviating eigenvalues $\lambda_2$ to $\lambda_5$ after removing the market effect.}
\label{Fig:AgriFut_ulargest}
\end{figure}

After removing the market effect contained in the correlation matrix $\mathbf{C}$, we next focus on the remaining largest eigenvalues and their corresponding eigenvectors. It is found that there are modular structures among the significant participants of these eigenvectors, implying the sector-like effect of other largest eigenvalues which deviate from RMT results. Figure~\ref{Fig:AgriFut_ulargest} shows all the components of the eigenvectors $\mathbf{u}^2$, $\mathbf{u}^3$, $\mathbf{u}^4$ and $\mathbf{u}^5$ associated with the four largest deviating eigenvalues $\lambda_2$ to $\lambda_5$. For convenience, we select the part beyond the red dotted line of each eigenvector as its significant participants. Table~\ref{tab:significant components of the largest eigenvectors} lists the relevant information of agricultural futures corresponding to these components.

From Table~\ref{tab:significant components of the largest eigenvectors}, it can be found that in the case of the positive sign, the agricultural futures corresponding to the important participants of $\mathbf{u}^2$ are all from NCDEX, showing that its modular structure is based on the regional properties. In contrast, the agricultural futures corresponding to the significant participants of $\mathbf{u}^3$ and $\mathbf{u}^4$ respectively belong to wheat products and soybean products, which indicates that these two modular structures are based on the common feature of agricultural products. Besides that, the agricultural futures corresponding to the important participants of $\mathbf{u}^5$ are all from CME and all belong to dairy products, manifesting that its modular structure may be based on regional properties or agricultural products. In addition, in the case of the negative sign, the significant participants of $\mathbf{u}^3$ and $\mathbf{u}^4$ have similar components, that is, 80\% from ICE and 60\% from coffee, which indicates that their modular structures are based on both regional properties and agricultural products.

The above results show that other largest deviating eigenvalues of the correlation matrix $\mathbf{C}$ can be used to identify futures groups, and the agricultural futures belonging to the same exchange or based on the same commodity property are highly correlated. As agricultural futures in the same group have similar price behaviors, investors should take the sector-like effect into consideration when using agricultural futures for risk diversification.

\begin{table*}[!t]
  \centering
  \renewcommand\tabcolsep{17pt} 
  \caption{Some significant participants of the eigenvectors $\mathbf{u}^2$ to $\mathbf{u}^5$ after removing the market effect}
      {
    \begin{tabular}{cccccc}
    \toprule
    Eigenvector & Sign & Label & Futures & Exchange & Country \\
    \midrule    
    $\mathbf{u}^2$ & $+$ & 65 & Guar Gum Refined Splits & NCDEX & India \\
                   & $+$ & 67 & Castor Seed & NCDEX & India \\
                   & $+$ & 64 & Guarseed & NCDEX & India \\
                   & $+$ & 68 & Jeera & NCDEX & India \\         
                   & $+$ & 61 & Mustard Seed & NCDEX & India \\         
                   & $+$ & 69 & Coriander & NCDEX & India \\         
                   & $+$ & 59 & Soybean & NCDEX & India \\
                   & $+$ & 60 & Soy Oil & NCDEX & India \\
    \midrule 
    $\mathbf{u}^3$ & $+$ & 48 & Feed Wheat & ICE & USA \\         
                   & $+$ & 53 & Wheat & ICE & USA \\     
                   & $+$ & 44 & Milling Wheat & EUREX & Europe \\
                   & $-$ & 49 & Coffee NY & ICE & USA \\
                   & $-$ & 3 & Coffee Arabica & BMF & Brazil \\
                   & $-$ & 50 & Robusta Coffee & ICE & USA \\
                   & $-$ & 52 & Sugar No.11 & ICE & USA \\
                   & $-$ & 51 & White Sugar & ICE & USA \\
    \midrule 
    $\mathbf{u}^4$ & $+$ & 9 & Soybean & CBOT & USA \\         
                   & $+$ & 5 & Soybean Meal & CBOT & USA \\         
                   & $+$ & 4 & Soybean Oil & CBOT & USA \\         
                   & $+$ & 2 & Soybean & BMF & Brazil \\
                   & $-$ & 3 & Coffee Arabica & BMF & Brazil \\
                   & $-$ & 49 & Coffee NY & ICE & USA \\
                   & $-$ & 53 & Wheat & ICE & USA \\
                   & $-$ & 48 & Feed Wheat & ICE & USA \\
                   & $-$ & 50 & Robusta Coffee & ICE & USA \\
    \midrule 
    $\mathbf{u}^5$ & $+$ & 14 & Cash Settled Cheese & CME & USA \\   
                   & $+$ & 13 & Class \Rmnum{3} Milk & CME & USA \\         
                   & $+$ & 12 & Cash Settled Butter & CME & USA \\
                   & $+$ & 15 & Non Fat Dry Milk & CME & USA \\  
    \bottomrule
    \end{tabular}}%
  \label{tab:significant components of the largest eigenvectors}%
\end{table*}%

% 2021/11/17, 17:51

\subsubsection{Smallest eigenvalues and highly correlated futures pairs} 

Having uncovered the economic information contained in the deviating eigenvalues larger than $\lambda_{\max}^{\text{RMT}}$, we next focus on the minimum eigenvalues and their eigenvectors. According to \cite{Plerou-Gopikrishnan-Rosenow-Amaral-Guhr-Stanley-2002-PhysRevE}, the eigenvectors associated with the smallest eigenvalues comprise the pairs of stocks with highest correlations in the empirical correlation matrix of the US stock market. Similar results were obtained in the global crude oil market \citep{Dai-Xie-Jiang-Jiang-Zhou-2016-EmpirEcon}.

\begin{figure}[!t]
\centering
\includegraphics[width=0.324\linewidth]{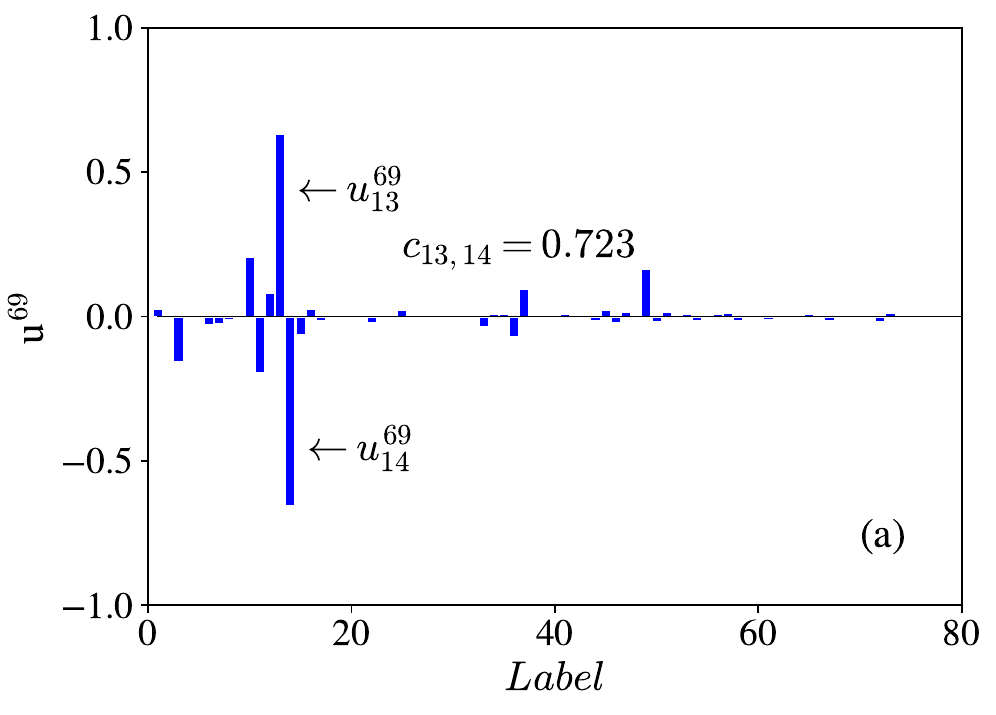}
\includegraphics[width=0.324\linewidth]{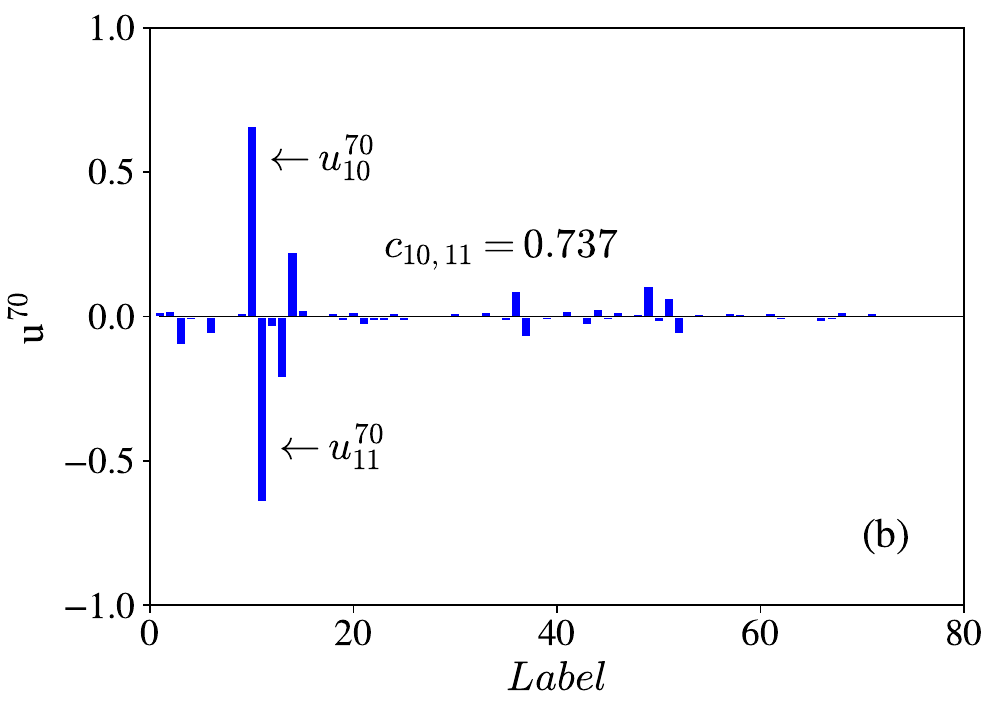}
\includegraphics[width=0.324\linewidth]{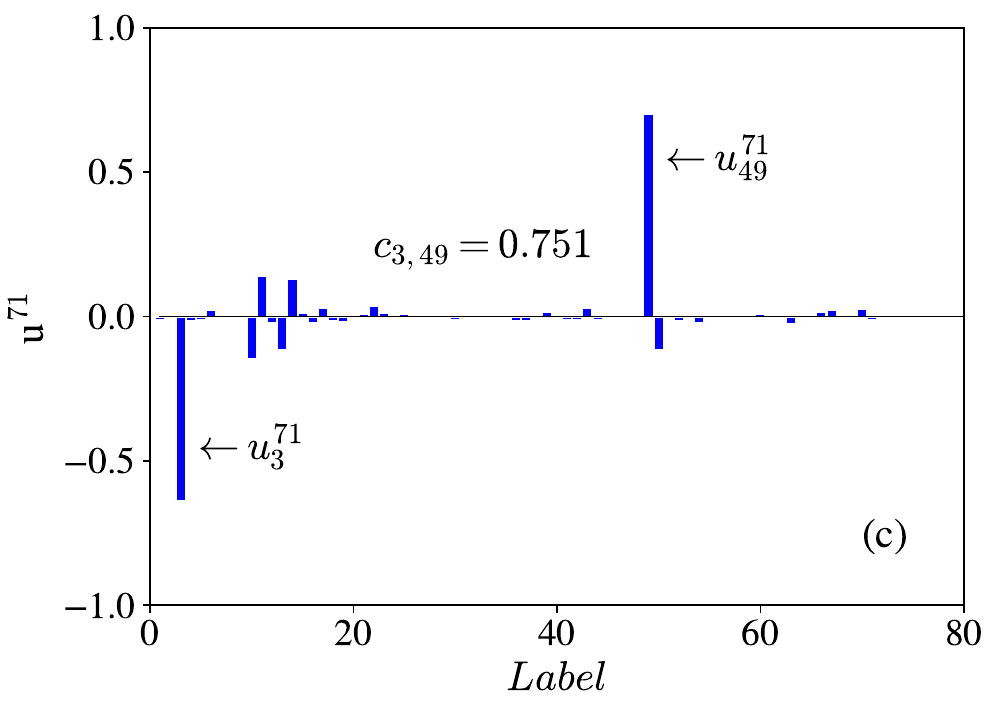}
\includegraphics[width=0.324\linewidth]{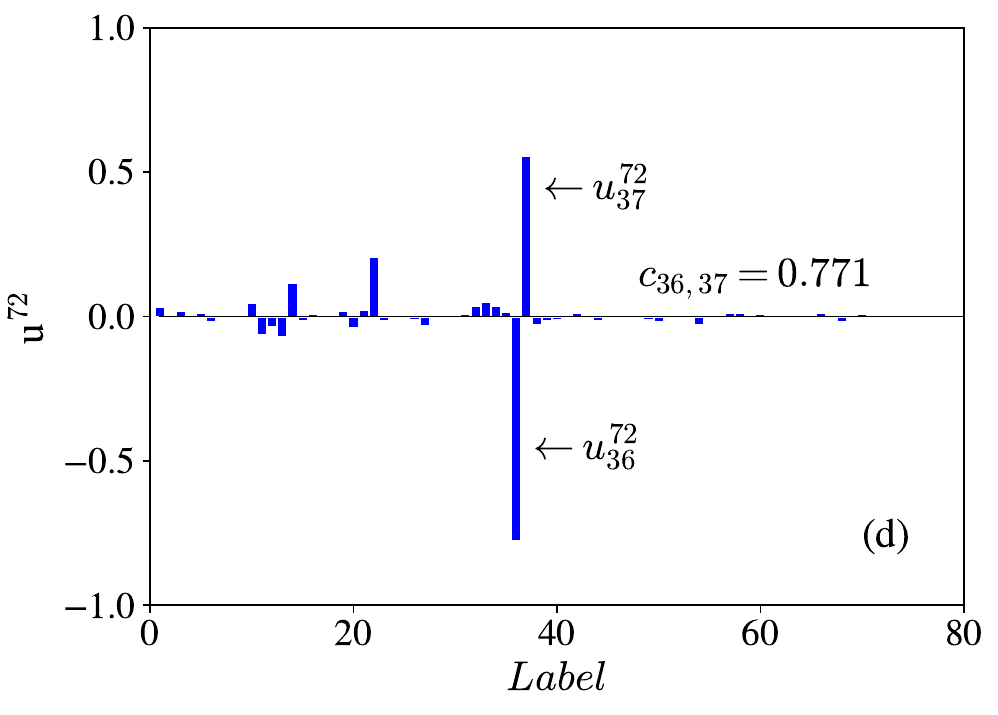}
\includegraphics[width=0.324\linewidth]{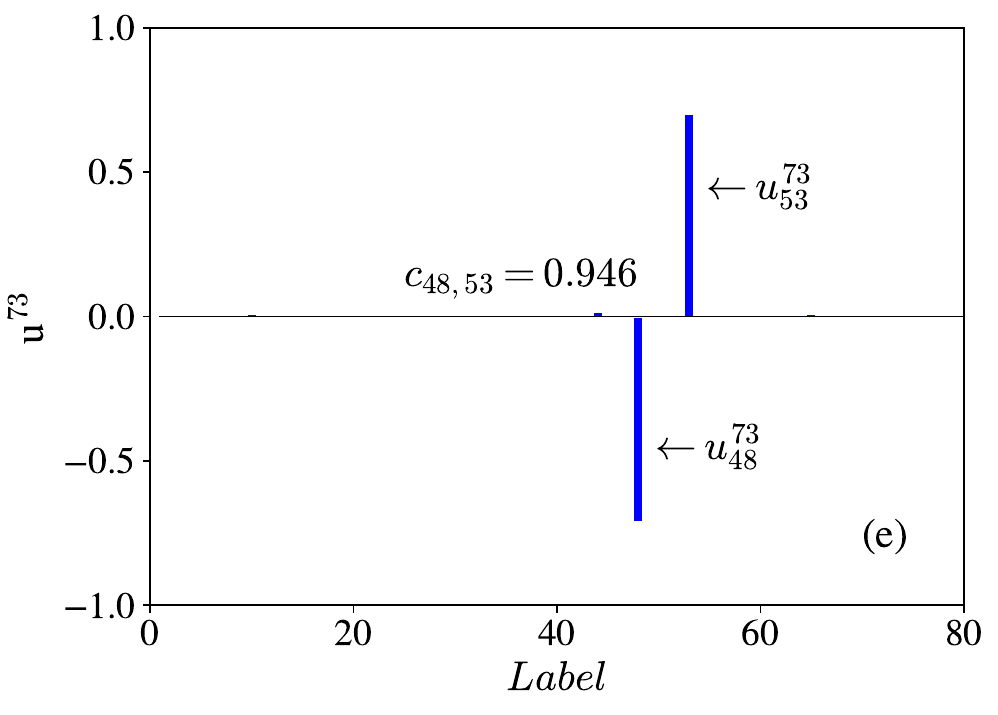}
\includegraphics[width=0.324\linewidth]{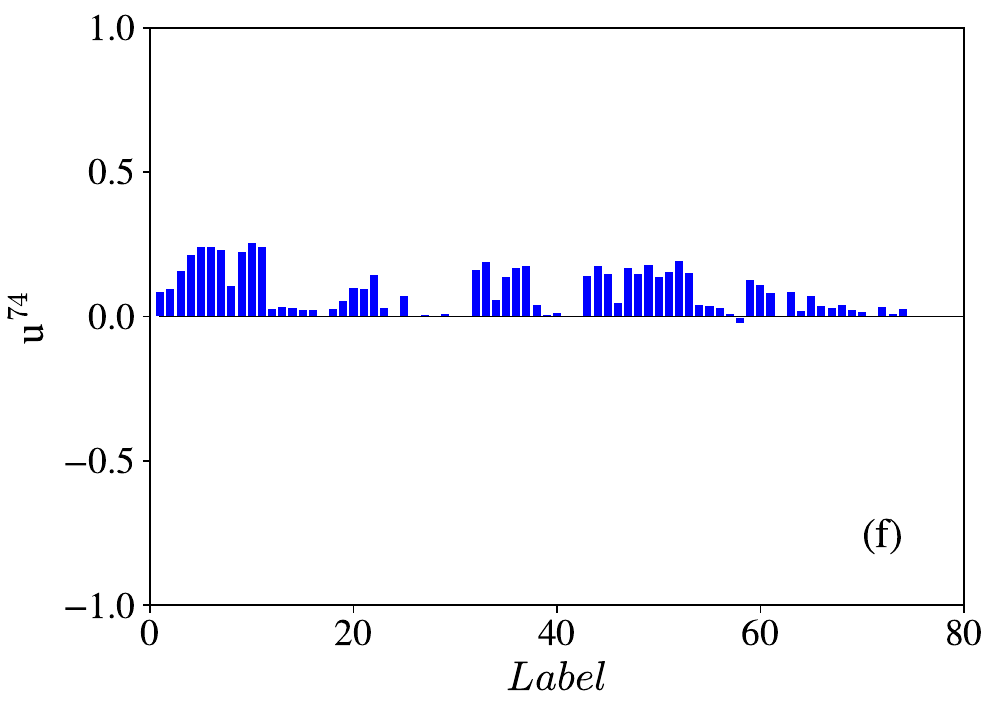}
\caption{Components of the eigenvectors $\mathbf{u}^{69}$ (a), $\mathbf{u}^{70}$ (b), $\mathbf{u}^{71}$ (c), $\mathbf{u}^{72}$ (d), $\mathbf{u}^{73}$ (e) and $\mathbf{u}^{74}$ (f) corresponding to the six smallest deviating eigenvalues $\lambda_{69}$ to $\lambda_{74}$ after removing the market effect.}
\label{Fig:AgriFut_usmallest}
\end{figure}

Figure~\ref{Fig:AgriFut_usmallest} shows all the components of the eigenvectors $\mathbf{u}^{69}$ to $\mathbf{u}^{74}$corresponding to the six smallest eigenvalues respectively after removing the market effect. Except for $\mathbf{u}^{74}$, we can observe a pair of components which are featured by opposite signs and significantly large values in each plot of $\mathbf{u}^{69}$, $\mathbf{u}^{70}$, $\mathbf{u}^{71}$, $\mathbf{u}^{72}$ and $\mathbf{u}^{73}$. We locate the labels of these components and compute the corresponding correlation coefficient between the return series of each pair. The two agricultural futures for $\mathbf{u}^{73}$ are the Feed Wheat Futures and Wheat Futures of ICE, whose correlation coefficient $c_{48,53}=0.946$ is the largest among all empirical correlation coefficients. The pair for $\mathbf{u}^{72}$ are the Soybean Oil Futures and RBD Palm Olein Futures of DCE, whose correlation coefficient $c_{36,37}=0.771$ is the second largest, and the pair for eigenvector $\mathbf{u}^{71}$ are BMF's Coffee Arabica Futures and ICE's Coffee NY Futures with $c_{3,49}=0.751$, which is the third largest among all correlation coefficients. The Wheat Futures and Red Hard Winter Wheat Futures of CBOT with $c_{10,11}=0.737$ are the fourth largest correlation coefficient for $\mathbf{u}^{70}$, and the fifth largest is $c_{13,14}=0.723$ between the Class \Rmnum{3} Milk Futures and Cash Settled Cheese Futures of CME for $\mathbf{u}^{69}$.

The above results show that the smallest deviating eigenvalues can represent the strongly correlated futures pairs in the correlation matrix, except for the minimum eigenvalue. Similar to the sector-like effect of the largest eigenvalues, these futures pairs are also based on the same exchange or similar agricultural properties. In addition, the price movements of these futures pairs are also highly consistent.

% 2021/11/17, 19:52

\section{Conclusions}
\label{S1:Conclude}

% \subsection{Conclusions} 

To summarize, we have clarified the correlation structure of the global agricultural futures market involving 74 agricultural futures traded on various exchanges all over the world by using the random matrix theory. Several statistical properties of the global agricultural futures market are revealed as follows.

We find that the empirical distribution of these correlation coefficients between various return series is asymmetric and right skewed, implying that positive correlations are more prevalent in the correlation structure of the agricultural futures market. By testing the eigenvalue statistics of the empirically measured correlation matrix $\mathbf{C}$ against a random matrix $\mathbf{R}$, we unveil that a considerable proportion of the eigenvalues of $\mathbf{C}$ fall outside the theoretical prediction range. Further, we investigate the deviations from RMT, and conclude that there exists abundant economic information in these deviating eigenvalues and their corresponding eigenvectors.

To be more specific, the largest eigenvalue reflects a collective market effect of the global agricultural futures market common to all agricultural futures. We construct eigenportfolios based on the eigenvectors of the correlation matrix and find that the returns of the eigenportfolio associated with the largest eigenvalue have an excellent linear relationship with the mean returns of the agricultural futures, while other eigenportfolios do not bear any significant correlations. Enlightened by this characteristic, we propose to construct an agricultural futures price index (AFPI) for the global agricultural futures market based on the eigenportfolio corresponding to the largest eigenvalue. The index AFPI is significantly greater than the average price series for most of the times, and shows a better performance than the latter under the buy-and-hold strategy.

% 2021/12/06, 15:34

After removing the collective market effect, other largest deviating eigenvalues of the correlation matrix exhibits the sector-like effect, and there are modular structures among the significant participants of their corresponding eigenvectors. These modular structures may be based on regional properties, or agricultural products, or both. Because the traders in the same exchange are the same, and the attributes of commodities themselves can reflect certain commonness, the agricultural futures in the same ``sector'' are highly correlated and have similar price behaviors. The elements of some eigenvector corresponding to these futures have higher magnitudes than other futures.

% 2021/12/06, 15:38

Except for the minimum eigenvalue, the other smallest deviating eigenvalues tend to have two large components in their eigenvectors, which represents pairs of return series with the highest correlation coefficients in the empirical correlation matrix. These futures pairs are mainly based on the same exchange or similar properties of agricultural products. Therefore, their price behaviors are highly consistent, with strong correlations.

% \subsection{Implications}

In the era of economic globalization, the degree of interdependence among economies in the world is deepening. Since 2020, the worldwide spread of COVID-19, combined with a more pessimistic outlook on the stability of agricultural systems, has increased the potential risks in the global agricultural futures market to some extent. Considering the great influence of agricultural futures market on spot market, all countries should be alert to the abnormal fluctuations of agricultural futures prices, so as to guarantee the effective supply of important agricultural products, and avoid causing food crisis. In addition, the maximum eigenvalue of the correlation matrix reflects the collective market effect, so the maximum eigenvalue can be regarded as one of the economic indicators to measure the systemic risk of the global agricultural futures market.

On the basis of standardizing the development of pre-existing trading varieties, commodity exchanges in various regions of the world should constantly promote variety innovation to enrich the varieties of agricultural futures, and continuously expand the breadth and depth of agricultural futures market, which can make significant contributions to reducing systemic risk in futures markets. By increasing the varieties of agricultural futures, the correlation of multiple sectors can be further reduced, and the differentiation between sectors can be realized, so as to provide more risk management tools for participants in the agricultural futures market. 

The agricultural futures in the same sector have certain commonness, showing similar price behavior and strong correlations between each other. Therefore, when investors use agricultural futures to hedge, they should take the sector-like effect into full consideration to manage investment risk and optimize asset allocation. Particularly, under the buy-and-hold strategy, the eigenportfolio based on the largest eigenvalue of the correlation matrix calculated by agricultural futures returns outperforms the equal-weight portfolio. Hence, when constructing investment portfolios, investors can give priority to utilizing the eigenportfolio associated with the largest eigenvalue of the correlation matrix, so as to obtain higher returns.

It is well established in financial literature that cross-asset relationships vary across time. The proposed and applied technique in this paper is static, which does not account for temporal changes in the correlation structure. In future research, bringing a dynamic perspective to the evolution of the correlation structure can be considered as a promising issue for extension. Furthermore, based on the primary work of this research, more light can be shed on quantifying the systemic risk in the global agricultural futures market and measuring the risk spillover between agricultural futures market and other financial markets.

% 2021/12/06, 21:53

\section*{Acknowledgment}

% % This work was supported by ...
This work was partly supported by the National Natural Science Foundation of China (72171083), the Shanghai Outstanding Academic Leaders Plan, and the Fundamental Research Funds for the Central Universities.

\section*{Data availability}

Agricultural futures data sets related to this article can be found at the Wind (https://www.wind.com.cn) and Bloomberg (https://www.bloomberg.com) databases.

%the National Natural Science Foundation of China under grants U1811462 
%\newpage
%
%\bibliography{E:/papers/Auxiliary/Bibliography}
%\bibliography{E:/Auxiliary/Bibliography}
%\bibliography{BibITN,BibRCE}%Bibliography,BibRCE,
%\bibliographystyle{plain}

% \end{CJK*}
\end{document}